\documentclass{article}

\usepackage{arxiv}

\usepackage[utf8]{inputenc} 
\usepackage[T1]{fontenc}    
\usepackage{hyperref}       
\usepackage{url}            
\usepackage{booktabs}       
\usepackage{amsfonts}       
\usepackage{nicefrac}       
\usepackage{microtype}      
\usepackage{graphicx}
\usepackage{natbib}
\usepackage{doi}
\usepackage{tabu}                      
\usepackage{booktabs}                  
\usepackage{lipsum}                    
\usepackage{mwe}                       
\usepackage{float}
\usepackage{multirow}
\usepackage{pifont}
\usepackage{subcaption}
\usepackage{xcolor}

\title{Towards the Target and Not Beyond:\\
2D vs 3D Visual Aids in MR-Based Neurosurgical Simulation}

\author{
  \href{https://orcid.org/0000-0002-1475-2751}{\includegraphics[scale=0.06]{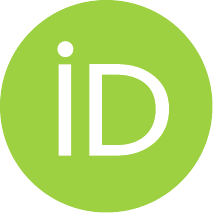}\hspace{1mm}Pasquale Cascarano} \\
  Dept. of the Arts\\
  University of Bologna\\
  Bologna, Italy \\
  \texttt{pasquale.cascarano2@unibo.it}
  \And
  \href{https://orcid.org/0009-0004-1667-6707}{\includegraphics[scale=0.06]{orcid.pdf}\hspace{1mm}Andrea Loretti} \\
  Dept. of the Arts\\
  University of Bologna\\
  Bologna, Italy \\
  \texttt{a.loretti@unibo.it}
  \And
  \href{https://orcid.org/0000-0002-9394-5841}{\includegraphics[scale=0.06]{orcid.pdf}\hspace{1mm}Matteo Martinoni} \\
  Dept. of Neurosurgery\\
  IRCCS Institute of Neurological Sciences\\
  Bologna, Italy \\
  \texttt{m.martinoni@isnb.it}
  \And
  \href{https://orcid.org/0009-0007-6195-1706}{\includegraphics[scale=0.06]{orcid.pdf}\hspace{1mm}Luca Zanuttini} \\
  Dept. of Neurosurgery\\
  IRCCS Institute of Neurological Sciences\\
  Bologna, Italy \\
  \texttt{luca.zanuttini@studio.unibo.it}
  \And
  \href{https://orcid.org/0009-0002-6250-9737}{\includegraphics[scale=0.06]{orcid.pdf}\hspace{1mm}Alessio Di Pasquale} \\
  Dept. of the Arts\\
  University of Bologna\\
  Bologna, Italy \\
  \texttt{alessio.dipasquale@studio.unibo.it}
  \And
  \href{https://orcid.org/0000-0003-3058-8004}{\includegraphics[scale=0.06]{orcid.pdf}\hspace{1mm}Gustavo Marfia} \\
  Dept. of the Arts\\
  University of Bologna\\
  Bologna, Italy \\
  \texttt{gustavo.marfia@unibo.it}
}

\renewcommand{\shorttitle}{TOWARDS THE TARGET AND NOT BEYOND}

\hypersetup{
pdftitle={TOWARDS THE TARGET AND NOT BEYOND: 2D VS 3D VISUAL AIDS IN MR-BASED NEUROSURGICAL SIMULATION},
pdfsubject={Mixed Reality in medicine},
pdfauthor={},
pdfkeywords={Mixed Reality, Neurosurgical Training, External Ventricular Drainage, Catherer Placement, Visual Aids, Skill Retention.},
}

\begin{document}
\maketitle

\begin{figure}[H]
  \centering
  \includegraphics[width=\linewidth, alt={Neurosurgeon performing an operation in Mixed Reality.}]{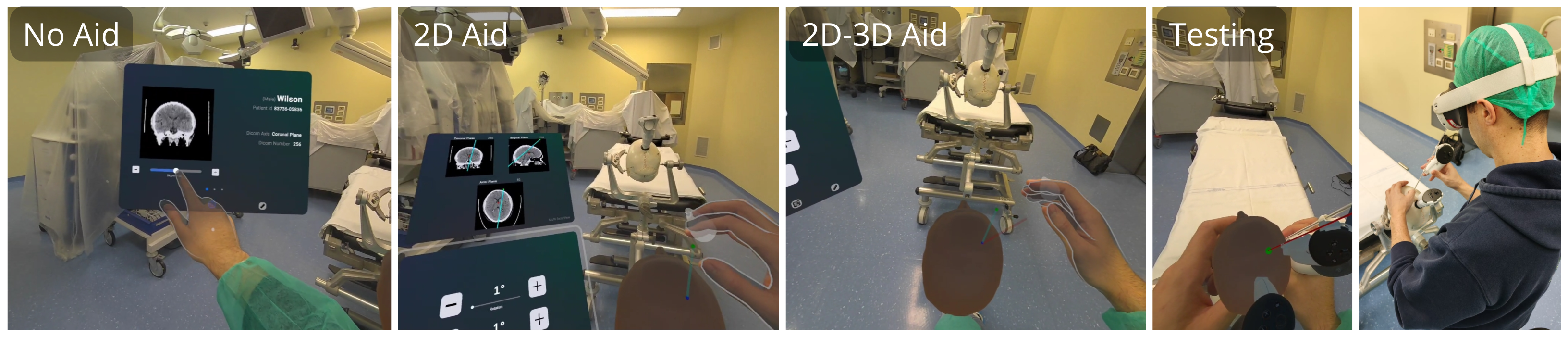}
  \caption{NEUROsurgical MIXed reality simulator (NeuroMix). The architecture consists of a Training system and a Testing system. Three training modalities with varying visual aids are compared: No Aid; 2D Aid; and 2D-3D Aid. During the Testing Phase a participant inserts a tracked Physical Catheter into a tracked Physical Skull in an unaided setting to assess skill retention.}
  \label{fig:teaser}
\end{figure}

\begin{abstract}
Neurosurgery increasingly uses Mixed Reality (MR) technologies for intraoperative assistance. The greatest challenge in this area is mentally reconstructing complex 3D anatomical structures from 2D slices with millimetric precision, which is required in procedures like External Ventricular Drain (EVD) placement.
MR technologies have shown great potential in improving surgical performance, however, their limited availability in clinical settings underscores the need for training systems that foster skill retention in unaided conditions.
In this paper, we introduce NeuroMix, an MR-based simulator for EVD placement. 
We conduct a study with 48 participants to assess the impact of 2D and 3D visual aids on usability, cognitive load, technology acceptance, and procedure precision and execution time. 
Three training modalities are compared: one without visual aids, one with 2D aids only, and one combining both 2D and 3D aids. 
The training phase takes place entirely on digital objects, followed by a freehand EVD placement testing phase performed with a physical catherer and a physical phantom without MR aids. 
We then compare the participants’ performance with that of a control group that does not undergo training.
Our findings show that participants trained with both 2D and 3D aids achieve a 44\% improvement in precision during unaided testing compared to the control group, substantially higher than the improvement observed in the other groups. All three training modalities receive high usability and technology acceptance ratings, with significant equivalence across groups. The combination of 2D and 3D visual aids does not significantly increase cognitive workload, though it leads to longer operation times during freehand testing compared to the control group.
\end{abstract}


\section{Introduction}
Augmented Reality (AR) and Mixed Reality (MR) technologies are revolutionizing medical education and intraoperative assistance \cite{giraldo2025advances}. 
Application of experiential training for medical students and residents entails the use of immersive and interactive simulators with no need for cadaver practice, thus avoiding inherent ethical concerns \cite{james2020use}. 
During surgery, AR/MR systems overlay virtual anatomical information or guidance visual cues onto the real operative field, allowing surgeons to visualize target structures and optimal trajectories in situ~\cite{colombo_mixed_2023,kazemzadeh_advances_2023}. 
These intuitive and real-time visual aids have the potential to enhance intraoperative navigation and reduce the cognitive load of correlating imaging with patient anatomy~\cite{isikay_narrative_2024}, and to improve surgical accuracy \cite{lungu2021review}.

External Ventricular Drain (EVD) placement (or ventriculostomy) is a critical neurosurgical procedure that stands to benefit from AR/MR technologies. 
The procedure involves inserting a catheter into the brain’s ventricular system for CerebroSpinal Fluid (CSF) drainage, to relieve intracranial pressure\cite{greenberg_anatomy_2023}. 
The surgeon identifies the correct entry point on the skull (commonly Kocher’s point), performs a burr hole craniotomy, opens the dura mater, and carefully advances the catheter through brain tissue into the lateral ventricle (commonly the Monro's foramen is the target point), ensuring proper depth and trajectory to avoid complications.
This procedural task is typically performed freehand at the bedside, using anatomical landmarks to identify the ventricular targets \cite{greenberg_anatomy_2023,flint2013simple}. 
However, this blind technique has some limitations, e.g. catheter misplacement occurs in up to 60\% of cases~\cite{alizadeh_virtual_2024}. 
These misplacements result in inadequate CSF drainage, but may also lead to other complications, potentially necessitating multiple repositioning attempts, each carrying added risk. 
The high error rate of the freehand approach underscores the clinical need for better guidance tools in ventriculostomy. 
A possible alternative to freehand catherer placement, already employed in clinical practice, is the use of neuronavigation systems \cite{gumprecht1999brainlab,alazri2017placement}. 
These systems rely on the real-time alignment of preoperative CT or MRI scans with the patient’s physical anatomy via fiducial markers and/or surface registration. 
The surgical instruments are then tracked (typically via optical or electromagnetic sensors), allowing the clinicians to visualize the catheter’s position and trajectory on a monitor.
The literature shows that when residents perform EVD placement under neuronavigation guidance, they achieve significantly higher accuracy than the conventional freehand approach~\cite{alazri2017placement}. 
Despite their effectiveness, these systems are often costly, require substantial setup time, are not always available in emergency contexts and there is a potential risk that overreliance on such technology may hinder the development of skill retention \cite{alazri2017placement}.

This underscores the need for training systems that enhance performance as well as skill retention.
To this end, AR/MR technologies have proven to be highly effective in a range of training-intensive domains, including procedural operations, technical maintenance, and complex assembly tasks\cite{kaplan2021effects,daling2024effects,borsci2015empirical}. 
A key characteristic of procedural tasks in the medical domain lies in the nature of the visual data involved. 
For example, unlike operators engaged in assembly or maintenance tasks, where 2D representations are typically schematic, symbolic, and standardized~\cite{agrawala2003designing,katsioloudis2014comparative}, medical professionals must interpret Computed Tomography (CT) scans \cite{kazemzadeh_advances_2023,durrani_virtual_2022} that portray complex, patient-specific internal anatomy. 
These 2D images are not abstract representations, but real cross-sections of anatomical structures that vary from individual to individual \cite{krabbe1998circle}.
Accordingly, clinicians are required to conduct complex spatial reasoning, combine information across various planes, and make precise judgments in conditions of uncertainty and clinical risk \cite{sidhu2004interpretation,yohannan2024visualization}.

This complexity makes the design of medical AR/MR training interfaces especially crucial: they must support intuitive spatial reasoning, reinforce procedural steps, and promote skill retention for accurate performances in unaided scenarios.
To this scope, visual 2D and 3D aids, such as contextual overlays, annotations, ghosted objects, and animated instructions \cite{lin2021labeling,hajahmadi2024investigating}, play a crucial role in enhancing user understanding and task execution \cite{kruger2022learning,tang2020evaluating}.

The present study evaluates NeuroMix, an MR simulator for EVD placement, involving medical residents with basic neuroimaging knowledge but no prior hands-on experience with the procedure.
We compare three distinct learning modalities, namely No Aid, 2D Aid, and 2D–3D Aid, where users perform the EVD procedure virtually.
In the No Aid mode, no visual assistance is provided beyond the standard CT scans. 
The 2D Aid mode is inspired by neuronavigation interface \cite{gumprecht1999brainlab,alazri2017placement}. Axial, coronal, and sagittal CT slices are shown to the user, in addition to the real-time projection of a virtual catheter overlying the imaging planes to support trajectory planning. 
The 2D–3D Aid modality builds on the previous one by adding a 3D animated guide that shows the correct placement of the virtual catherer in space.
Following the training session, all participants complete a testing phase where they are asked to perform the EVD procedure using a real catheter on a phantom skull, without access to MR-based visual guidance. 
A separate Control group, which does not undergo the training, completes only the testing phase allowing for comparative assessment of performance and skill retention across conditions. 
A pilot study with 10 participants was conducted before the full-scale
experiment to validate the system and address the challenges. The follow-up study involved 48 medical residents to address the following Research Questions (RQs): 

RQ1: How do users' perceived usability (measured by SUS), workload (measured by NASA-TLX), and technology acceptance (measured by TAM) differ when comparing the No Aid, 2D Aid, and 2D-3D Aid groups in EVD placement during the training phase?

RQ2: How do users' procedure precision and execution time differ when comparing the No Aid, 2D Aid, and 2D-3D Aid groups in EVD placement during the training phase?

RQ3: What is the impact of the No Aid, 2D Aid, and 2D-3D learning modalities for EVD placement on users' skill retention, measured by procedure accuracy and execution time, once these aids are no longer provided? 

This research assesses the usability, cognitive impact, and effectiveness of different training modalities in maximizing performance and short-term skill retention without aids. 
The goal is to identify which interface best promotes learning and independent execution, guiding the design of effective clinical MR simulators.

\section{Related Work}

AR/MR systems for EVD placement have been designed primarily to support intraoperative performance \cite{buwaider_augmented_2024}. 
In most experimental studies \cite{van_gestel_effect_2021,eom_accuracy_2024,li_wearable_2019,grunert_imaginer_2024,vychopen_imaginer_2025,eom_augmented_2024}, systems deployed have been evaluated mainly in terms of the accuracy of practitioners when supported by the AR/MR interface via 2D and 3D visual aids.
The pioneering works \cite{yudkowsky_practice_2013,hooten_mixed_2014} presented the first EVD guidance systems, using haptic feedback, varied virtual brains to simulate different realistic clinical scenarios and 3D overlays displayed on an external monitor. 
Similarly, in \cite{li_wearable_2019} the authors introduce one the first MR system for EVD placement with reduced catheter misplacement and number of insertion attempts in bedside procedures.
Later studies \cite{van_gestel_effect_2021,schneider2023augmented} found that AR increased placement accuracy and helps novices reach expert-level performance. 

The role played by visual aids used in these AR/MR systems is crucial  to provide contextual and real-time guidance regarding catheter trajectory, insertion angle, and distance to the target. 
In \cite{dominguezvelasco_augmented_2023,benmahdjoub_evaluation_2023, eom_augmented_2024,skyrman_augmented_2021} the authors proposed systems visualizing the ideal catheter trajectory on a physical skull model, along with virtual target point markers (Kocher’s point and/or Monro's foramen) or segmented anatomical structures such as ventricles. Similarly, some systems \cite{skyrman_augmented_2021,dominguezvelasco_augmented_2023,grunert_imaginer_2024} provided diagnostic images, textual instructions, and feedback panels to support the user during the EVD placement.
For example, \cite{eom2024did} proposed a system that tracks user gaze and triggers visual cues when key interface areas are overlooked.
In \cite{benmahdjoub_evaluation_2023}, different AR visualization modes were evaluated for EVD procedure: 3D aids resulted to be more accurate, intuitive, and cognitively efficient, whereas 2D aids provided useful support for setting catherer's insertion depth and angles.

Most of the AR/MR systems developed for EVD placement rely on head-mounted displays, primarily the Microsoft HoloLens \cite{eom_augmented_2024,umana2021multimodal}.
To align virtual content with the patient’s anatomy, systems adopted a combination of registration methods and tracking technologies. 
Registration is typically performed either manually, by visually aligning holograms to anatomical landmarks  \cite{eom_accuracy_2024,eom_augmented_2024}, or via fiducial markers placed on the scalp to enable more precise and repeatable alignment \cite{li_wearable_2019,skyrman_augmented_2021,schneider2021augmented}.
For spatial tracking, systems employ either outside-in tracking with external optical devices \cite{skyrman_augmented_2021,eom_augmented_2024}, or inside-out tracking, which uses the headset’s built-in sensors for autonomous localization without external references \cite{van_gestel_effect_2021,grunert_imaginer_2024}. 
A possible alternative for both registration and real-time tracking is also represented by computer vision techniques \cite{robertson2021frameless}. 

Validation of these systems has been conducted on phantom heads \cite{van_gestel_effect_2021, eom_accuracy_2024}, 3D-printed skulls \cite{dominguezvelasco_augmented_2023}, cadaver specimens \cite{grunert_imaginer_2024}, and real patients \cite{li_wearable_2019}. 
Evaluated outcomes included target deviation from the planned trajectory \cite{li_wearable_2019,skyrman_augmented_2021}, number of catheter passes \cite{eom_accuracy_2024}, procedural time \cite{dominguezvelasco_augmented_2023}, and trajectory alignment error \cite{benmahdjoub_evaluation_2023,grunert_imaginer_2024}. 
Some studies also reported subjective metrics, such as user confidence and perceived spatial awareness \cite{eom2024did,benmahdjoub_evaluation_2023}. 

In contrast to existing literature, this paper offers new contributions to system design and methodological analysis.
The system is constructed on the Meta Quest 3, utilizing its increased field of view, improved rendering, and reduced cost compared to the Microsoft HoloLens. 
By leveraging the headset's inside-out tracking, we developed a tracking system using the Quest controllers to offer a precise and portable simulator with no need for external trackers.
We developed and compared three different interface modalities, each incorporating varying levels of 2D and 3D visual aids.
Notably, we proposed a hybrid visualization, combining both 2D and 3D elements inspired by the interface design of clinical neuronavigation systems relying on CT image slices, which are familiar to users and aligned with their formal training. 
To the best of our knowledge, a similar interface concept was proposed only in \cite{skyrman_augmented_2021}. However, their system is non-immersive, with visual aids shown on a monitor rather than in the user's field of view. 
From a methodological standpoint, another key innovation of our study is the assessment of training impact on skill retention in an unaided setting. 
After completing the training, participants were asked to perform the task again without any visual support from the system. To the best of our knowledge, only one previous work \cite{dominguezvelasco_augmented_2023} has addressed skill retention, though it relied on monitor-based visual guidance. 
This highlights a gap in the current literature since many studies demonstrate the effectiveness of AR/MR systems in improving procedural performance \cite{eom_augmented_2024, benmahdjoub_evaluation_2023}, but they often do not assess whether those improvements persist once the visual aids are removed.

\begin{figure*}
    \centering
    \includegraphics[width=1\linewidth]{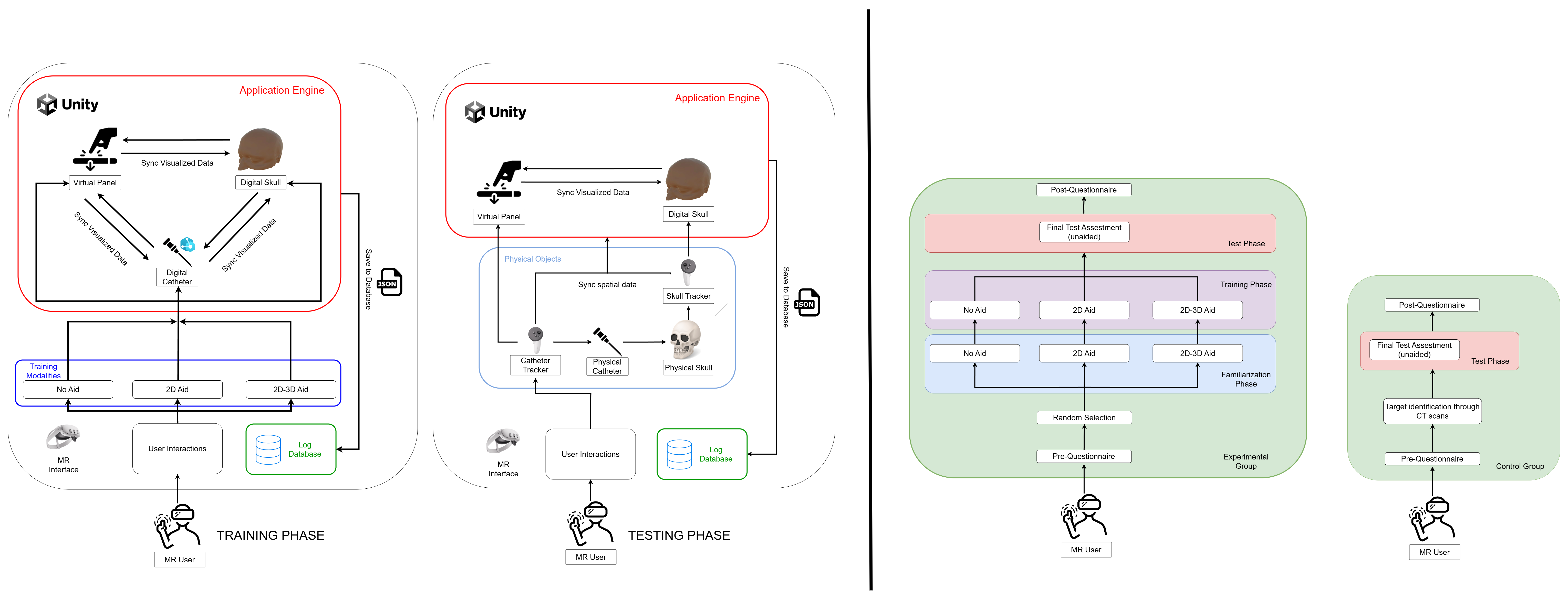}
    \caption{NeuroMix. The architectures used during the Training and Testing Phases (left panel) and the Experimental Setting (right panel).}
    \label{fig:System}
\end{figure*}

\section{Materials and Methods}

In this section we describe the training and testing systems, the initial pilot study for system refinement and the experimental design. 

\subsection{Training system design and architecture}

The architecture shown in \autoref{fig:System} illustrates the NeuroMix training system (left panel) which has been developed using Unity3D (v2022.3.10f1) and the Meta SDK (v62.0.0) \cite{MetaSDK2023}, and deployed on the Meta Quest 3 headset for UI interaction and controller tracking. 
The system architecture comprises three primary modules: the Application Engine, the Data Logger and the Training Modality Interface (see \autoref{fig:System}, left panel).

\subsubsection{The Application Engine and the Data Logger}

The Application Engine is responsible for visualizing and synchronizing data among three virtual components: the Virtual Panel, the Digital Skull, and the Digital Catheter.
These components are synchronized through a real-time, bidirectional synchronization system.

The Virtual Panel serves as an in-world interface that provides real-time feedback and data visualization to the user. 
The content displayed depends on the specific Training Modality Interface selected. 
In general, the Virtual Panel presents key information such as insertion depth, trajectory angles, and brain CT scans. 
The design is inspired by well-known software used in the medical setting, such as 3D Slicer and OsiriX \cite{rosset2004osirix,pieper20043d}, aiming to provide users with a familiar and intuitive interface \cite{salomoni2017diegetic}.
The Digital Skull represents the human skull of an anonymous patient.
This 3D digital element allows users to explore key anatomical landmarks, such as Kocher’s point, the standard neurosurgical entry site for EVD catheter insertion, and the Monro's foramen, which serves as the anatomical target for precise catheter placement \cite{greenberg_anatomy_2023}.
The Digital Skull was reconstructed using the 3D Slicer software \cite{pieper20043d} from a real brain CT dataset consisting of 512 sagittal, 512 coronal, and 128 axial images in DICOM format, each with a resolution of 512$\times$512 pixels.
The resulting 3D model was optimized in Blender \cite{mullen2011mastering} to ensure smooth performance and compatibility with the hardware constraints and calibrated in Unity to accurately align DICOM slices within the volume.
The Digital Catheter is the virtual surgical tool. A green sphere is placed at the tip of the Digital Catheter to enhance the user's manipulation. 

These components are integrated into an interactive workflow that guides the user through the key steps of the EVD placement procedure.
In the beginning, the system allows the user to position the Digital Skull comfortably in the environment, anchor it using a dedicated button, and choose one among three entry points highlighted on the Digital Skull. 
The user must identify and select the Kocher’s point. 
If the selected point is incorrect, the system provides error feedback on the Virtual Panel and requests a new selection. 
This encourage users to actively recognize anatomical landmarks, just as they would be required to do in a real surgical setting where the Kocher’s point is not pre-marked. 
Once the correct entry point is choosen, the virtual simulation of the EVD placement procedure begins. 
The user can define the Digital Catheter trajectory based on three degrees of freedom: Tilt, Rotation, and Depth.
The Digital Catheter is anchored at the selected point on the Digital Skull, and the user can manipulate the Tilt and Rotation components, either by directly adjusting it in the 3D space or by interacting with the Virtual Panel. 
However, Depth can only be chosen through the Virtual Panel sliders to allow millimeter-level precision \cite{mendes2019survey}.
After confirming the choices, the system evaluates the trial and provides the user with some post-trial feedback depending on the Training Modality Interface. To repeat the trial, the user can press the “Retry” button on the Virtual Panel. The system then restarts from the entry point selection phase, allowing iterative practice and progressive refinement of the insertion technique. 

Finally, the Data Logger module records user interactions logs during the experiment sessions.

\subsubsection{The Training Modality Interface}
NeuroMix provides three different training modalities: No Aid, 2D Aid and 2D-3D Aid.

\begin{figure}[!htb]
    \centering
    \begin{subfigure}[b]{0.23\linewidth} 
        \centering
        \includegraphics[width=\linewidth]{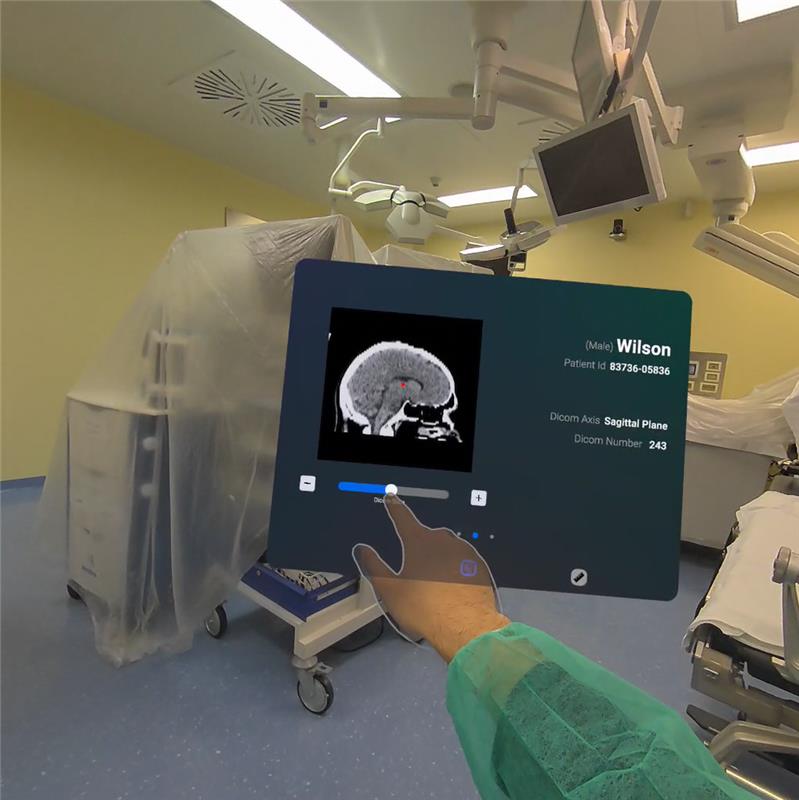}
        \caption{}
        \label{fig:NoAid1}
    \end{subfigure}
    \begin{subfigure}[b]{0.23\linewidth} 
        \centering
        \includegraphics[width=\linewidth]{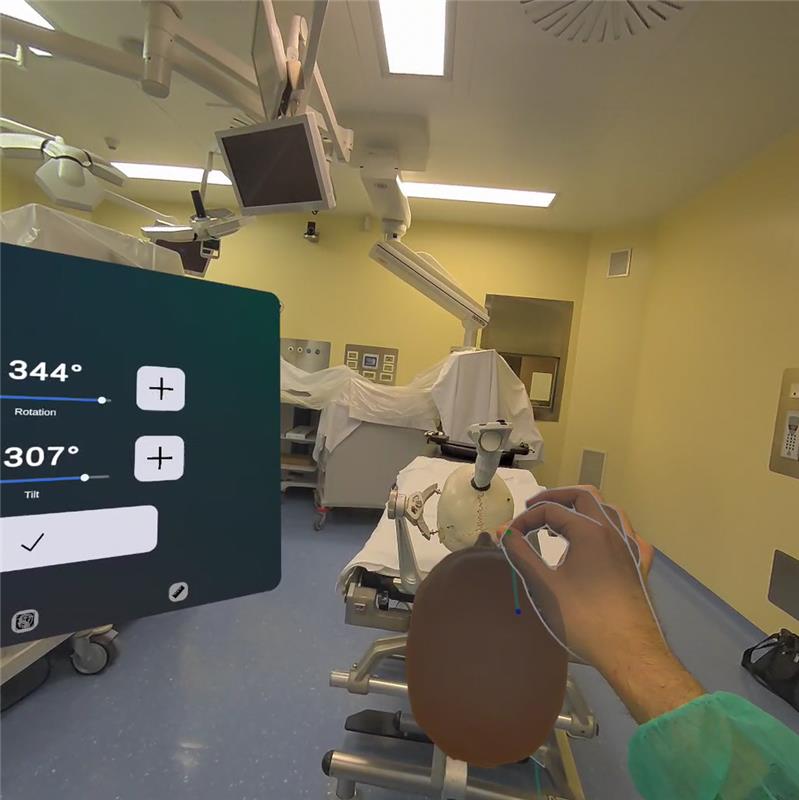}
        \caption{}
        \label{fig:NoAid2}
    \end{subfigure}
    \begin{subfigure}[b]{0.23\linewidth} 
        \centering
        \includegraphics[width=\linewidth]{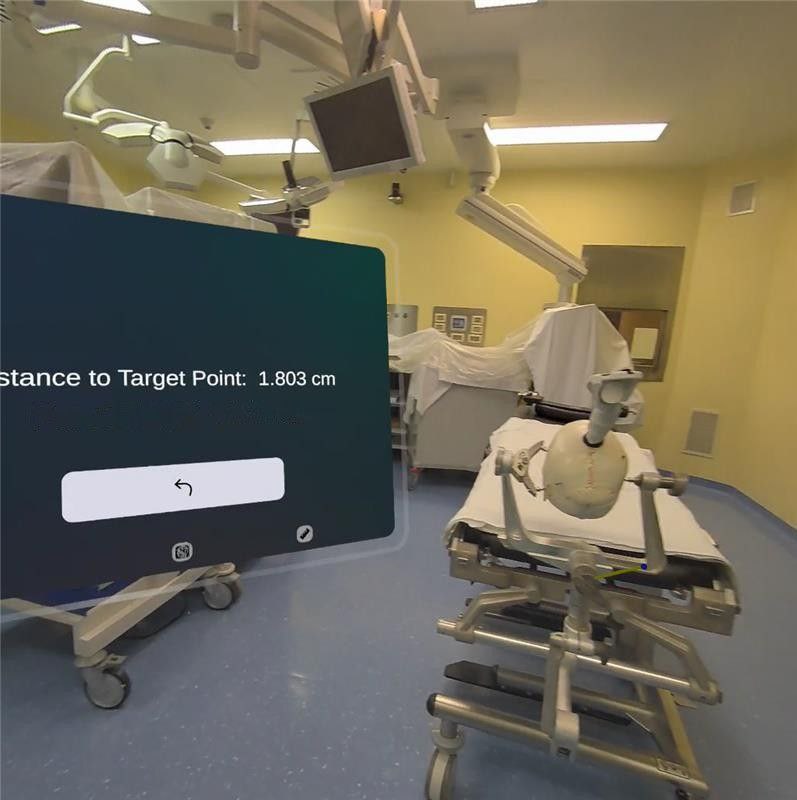}
        \caption{}
        \label{fig:NoAid3}
    \end{subfigure}
     \begin{subfigure}[b]{0.23\linewidth} 
        \centering
        \includegraphics[width=\linewidth]{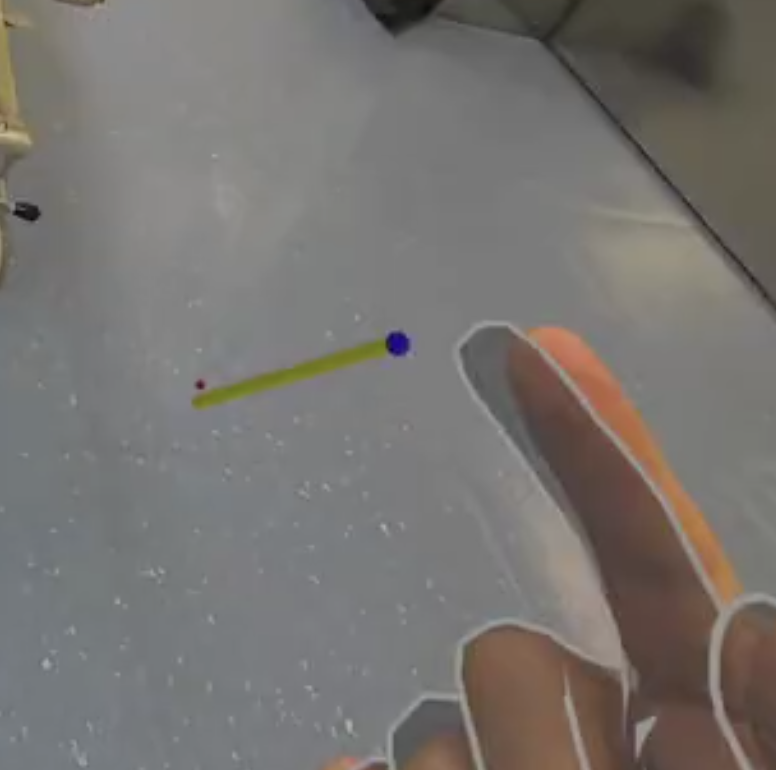}
        \caption{}
        \label{fig:NoAid4}
    \end{subfigure}
    \caption{No Aid modality interface.}
    \label{fig:NoAidAll}
\end{figure}

\paragraph{No Aid Modality} The user is provided only with CT images and the target point (Monro's foramen). 
All CT images are displayed on the Virtual Panel and the target point is highlighted with red dot. By interacting with the Virtual Panel, the user can move through the CT scans across all the anatomical planes (sagittal, axial, and coronal). 
The red dot is visible upon selecting the correct CT slice, requiring the user to actively locate the target (see \autoref{fig:NoAid1}).
No additional visual aids are provided during the virtual insertion. 
The user has the option to continue visualizing the CT scans on the Virtual Panel while selecting Tilt, Rotation and Depth values (see \autoref{fig:NoAid2}).
Upon confirming the selection, the Digital Catherer is anchored and the system provides different forms of feedback to the user. 
One is a textual feedback appearing on the Virtual Panel, namely the Euclidean distance (in centimeters) between the Digital Catheter's tip (operation point) and the digital anatomical target point (see \autoref{fig:NoAid3}).
The other is a 3D feedback: the Digital Skull is made transparent, allowing the user to visualize the anchored Digital Catherer redered yellow and the target point within the 3D environment, thus reinforcing spatial understanding and procedural precision (see \autoref{fig:NoAid4}). 

\begin{figure}[!htb]
    \centering
    \begin{subfigure}[b]{0.23\linewidth} 
        \centering
        \includegraphics[width=\linewidth]{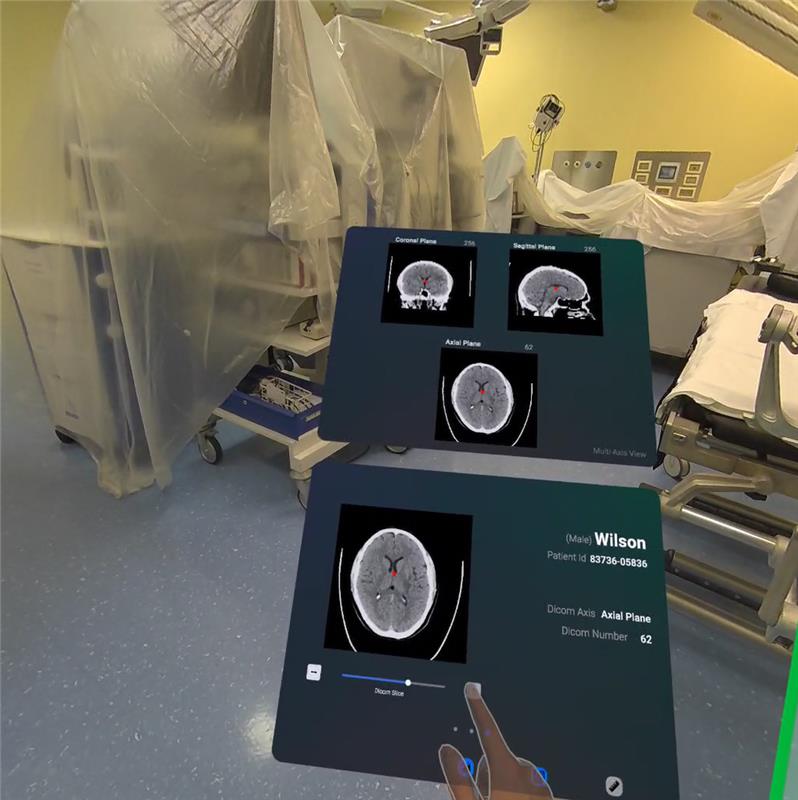}
        \caption{}
        \label{fig:2DAid1}
    \end{subfigure}
     \begin{subfigure}[b]{0.23\linewidth} 
        \centering
        \includegraphics[width=\linewidth]{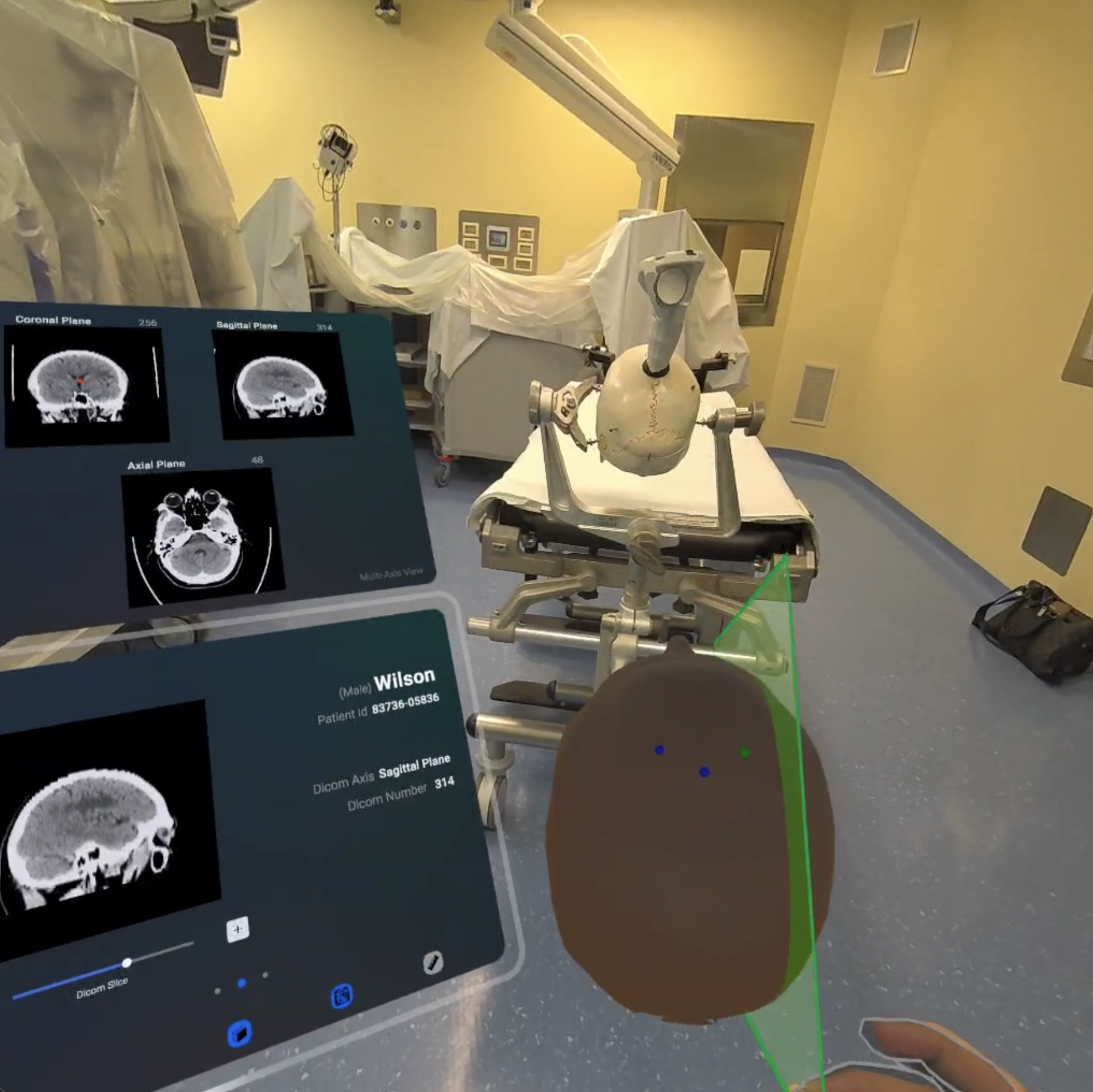}
        \caption{}
        \label{fig:2DAid2}
    \end{subfigure}
    \begin{subfigure}[b]{0.23\linewidth} 
        \centering
        \includegraphics[width=\linewidth]{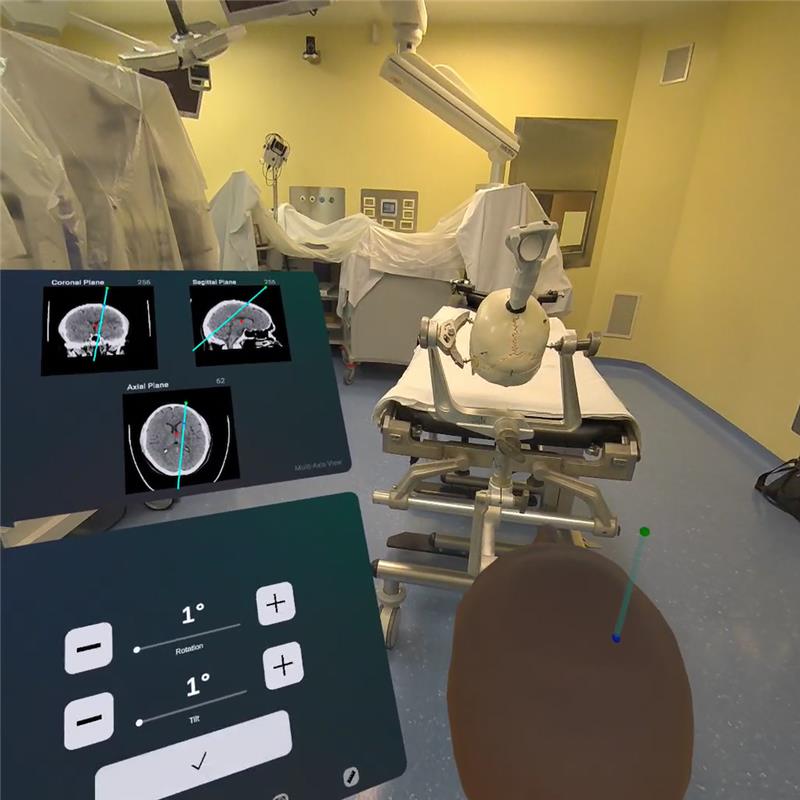}
        \caption{}
        \label{fig:2DAid3}
    \end{subfigure}
    \begin{subfigure}[b]{0.23\linewidth} 
        \centering
        \includegraphics[width=\linewidth]{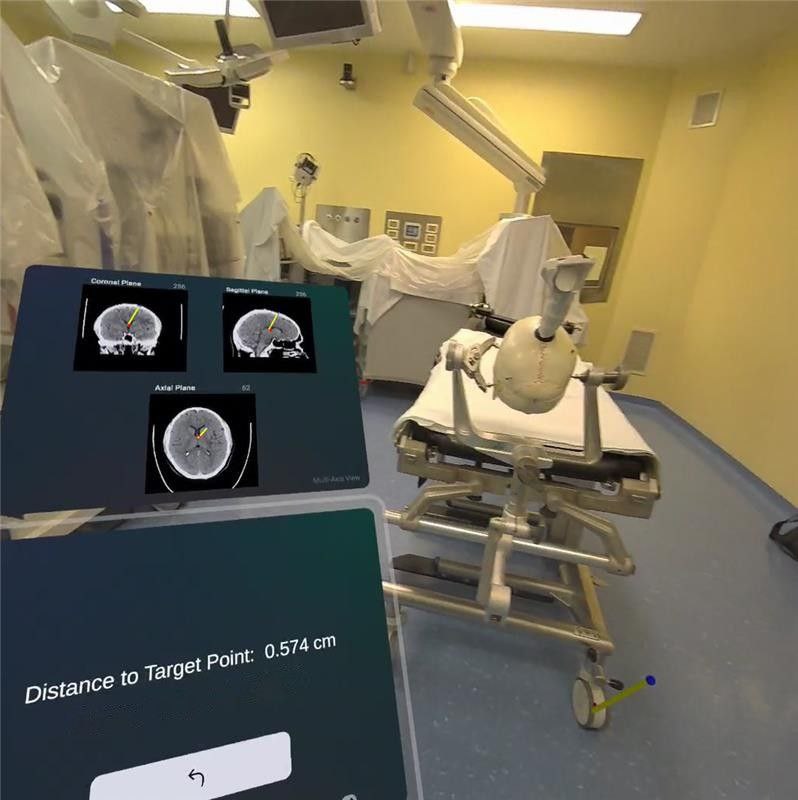}
        \caption{}
        \label{fig:2DAid4}
    \end{subfigure}
    \caption{2D Aid modality interface.}
    \label{fig:2DAidAll}
\end{figure}

\paragraph{2D Aid Modality} 
The interface is inspired by the layout of a neuronavigation system \cite{gumprecht1999brainlab,alazri2017placement}.
The Virtual Panel displays the anatomical planes simultaneously (see \autoref{fig:2DAid1}). 
To identify the target point highlighted with a red dot, the user can scroll the CT images using the buttons on the Virtual Panel or by dragging green virtual panels that appear on the Digital Skull (see \autoref{fig:2DAid2}).
This interface is designed to make the connection between the 2D CT slices on the Virtual Panel and the 3D Digital Skull more intuitive \cite{mendes2019survey}.
Once the red target point is identified, the user proceeds with the placement, and the selected CT slices are carried over as visual aids to the next steps. 
Then, the user selects the Rotation, Tilt, and Depth values.
However, this interface includes a real-time 2D projection of the Digital Catheter superimposed on the CT scan images. 
This projection is continuously updated while the user changes the Rotation, Tilt, and Depth values, providing 2D visual aids of the catheter’s orientation (see \autoref{fig:2DAid3}). 
After each attempt, the system provides the same feedback as the No Aid modality. In addition, the user is provided with the final catheter's 2D trajectories across the three anatomical planes on the CT scans (see \autoref{fig:2DAid4}).



\begin{figure}[!htb]
    \centering
    \begin{subfigure}[b]{0.23\linewidth} 
        \centering
        \includegraphics[width=\linewidth]{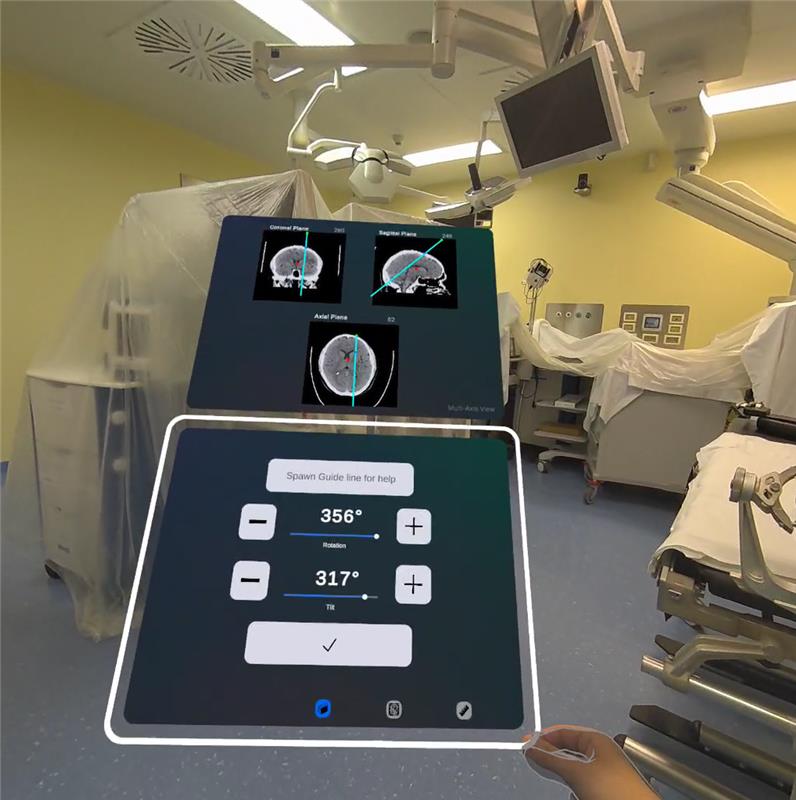}
        \caption{}
        \label{fig:3DAid1}
    \end{subfigure}
    \begin{subfigure}[b]{0.23\linewidth} 
        \centering
        \includegraphics[width=\linewidth]{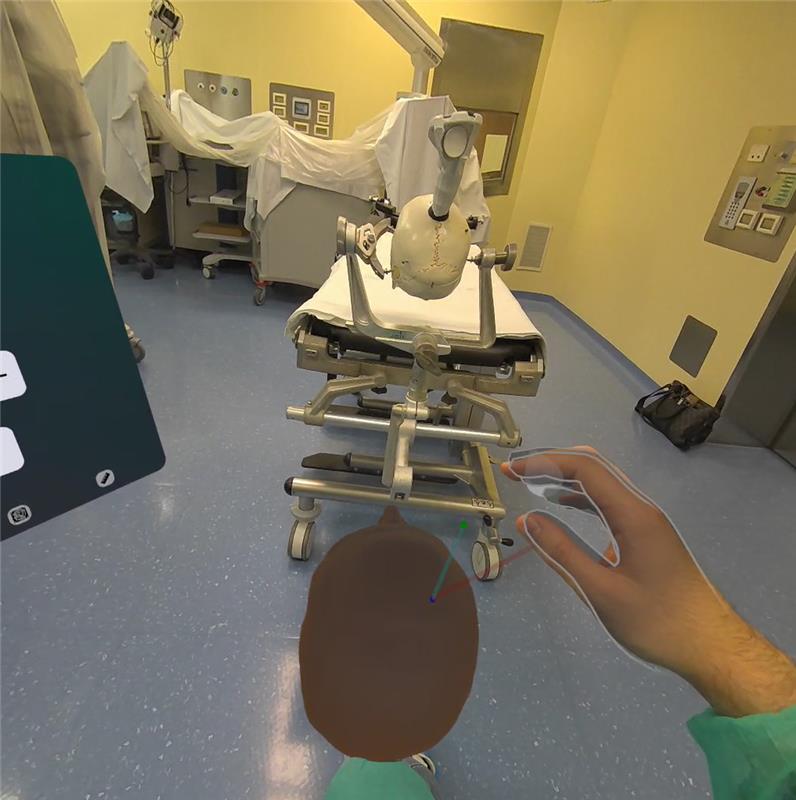}
        \caption{}
        \label{fig:3DAid2}
    \end{subfigure}
    \begin{subfigure}[b]{0.23\linewidth} 
        \centering
        \includegraphics[width=\linewidth]{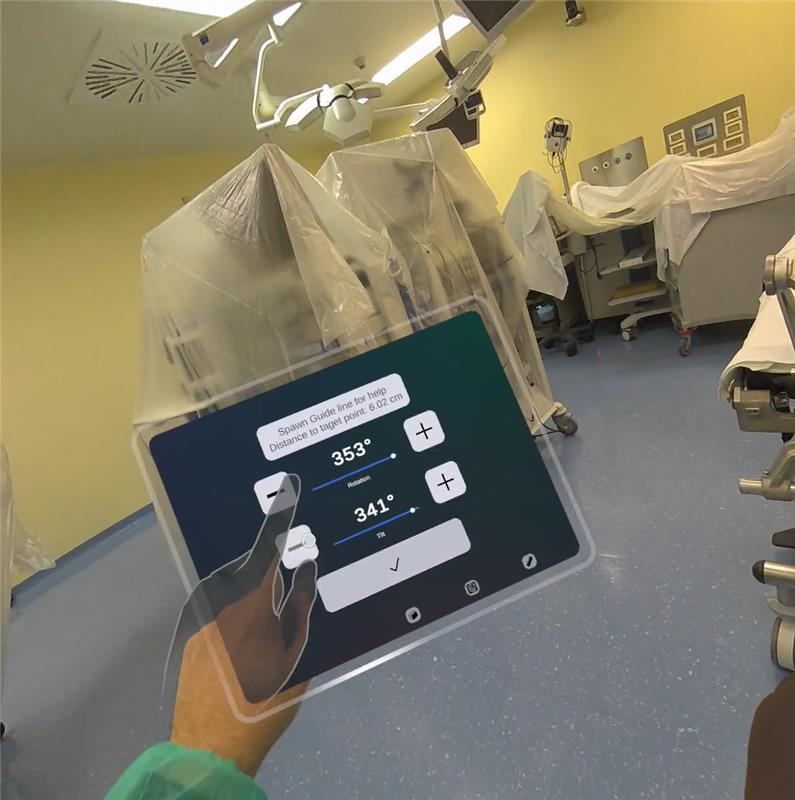}
        \caption{}
        \label{fig:3DAid3}
    \end{subfigure}
    \begin{subfigure}[b]{0.23\linewidth} 
        \centering
        \includegraphics[width=\linewidth]{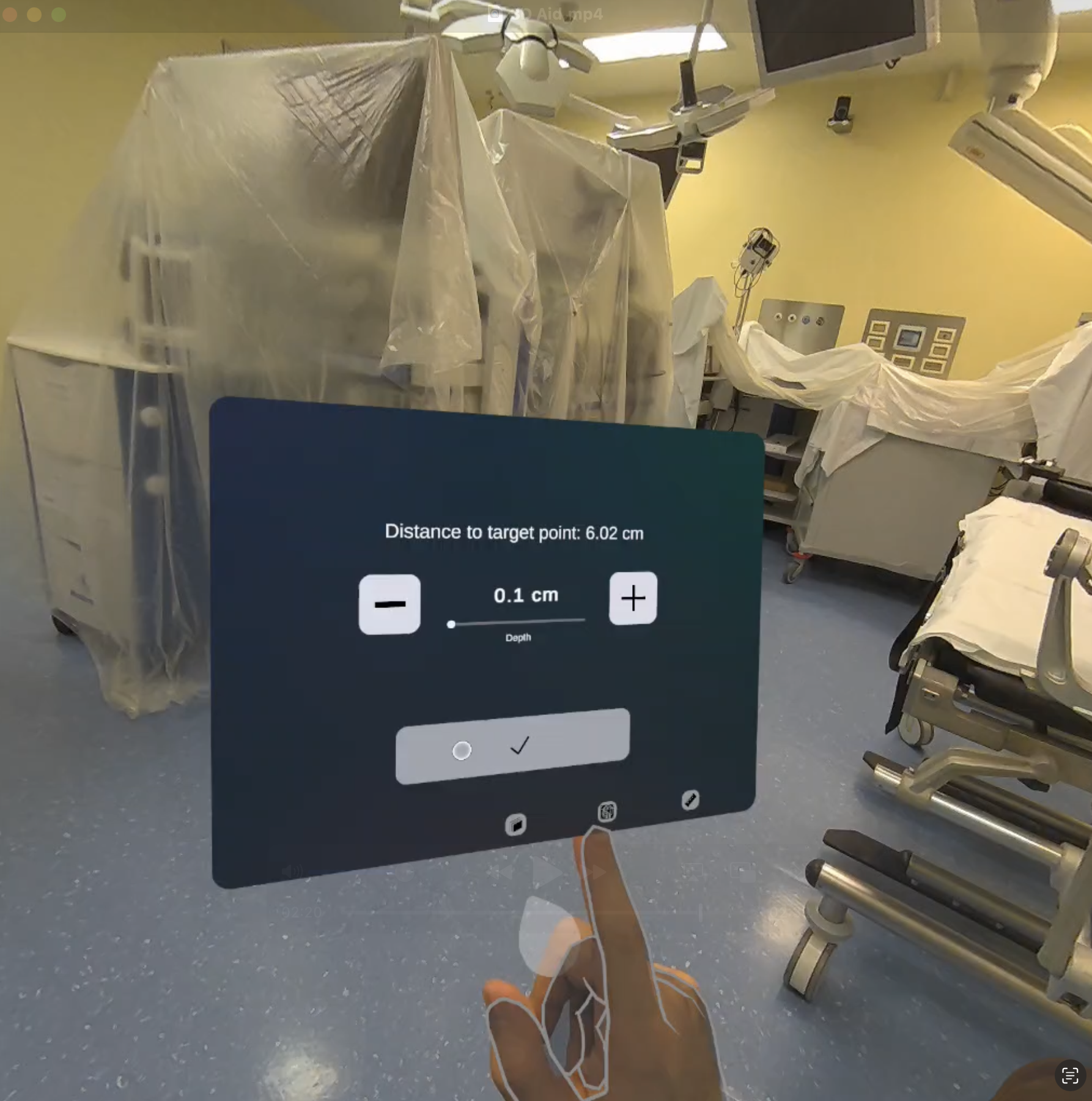}
        \caption{}
        \label{fig:3DAid4}
    \end{subfigure}
    \caption{2D-3D Aid modality interface.}
    \label{fig:3DAid}
\end{figure}

\paragraph{2D-3D Modality} This modality builds upon the 2D Aid modality (see \autoref{fig:3DAid1}). 
An additional visual aid is introduced in the form of a 3D trajectory guide to support the user in the selection of the Tilt and Rotation values. This red Guide Line visually indicates the optimal insertion path in 3D, including the correct angle. This allows participants to follow a predefined path that would lead to a "perfect" placement if executed accurately (see \autoref{fig:3DAid2}).
A dedicated button on the Virtual Panel allows the user to activate an animation. 
This animation is context-aware: starting from the Digital Catheter’s current orientation, it visually shows the movement required for the Digital Catheter to align with the ideal insertion angle (see \autoref{fig:3DAid3}).
The Depth component is then selected using the sliders of the Virtual Panel.   
Differently from the previous modalities, the Virtual Panel shows the user the ideal depth component value in order to reach the inner target point (see \autoref{fig:3DAid4}). 
After each virtual EVD placement, the system provides the same feedback as the 2D Aid modality. 

\subsection{Testing system design and architecture}

The architecture shown in \autoref{fig:System} illustrates the NeuroMix MR
testing system (left panel). 
The testing system is composed of three main components: the Physical Objects, the Application Engine, and the Data Logger.

\begin{figure}[!htb]
    \centering
    \begin{subfigure}[b]{0.24\linewidth} 
        \centering
        \includegraphics[width=0.815\linewidth]{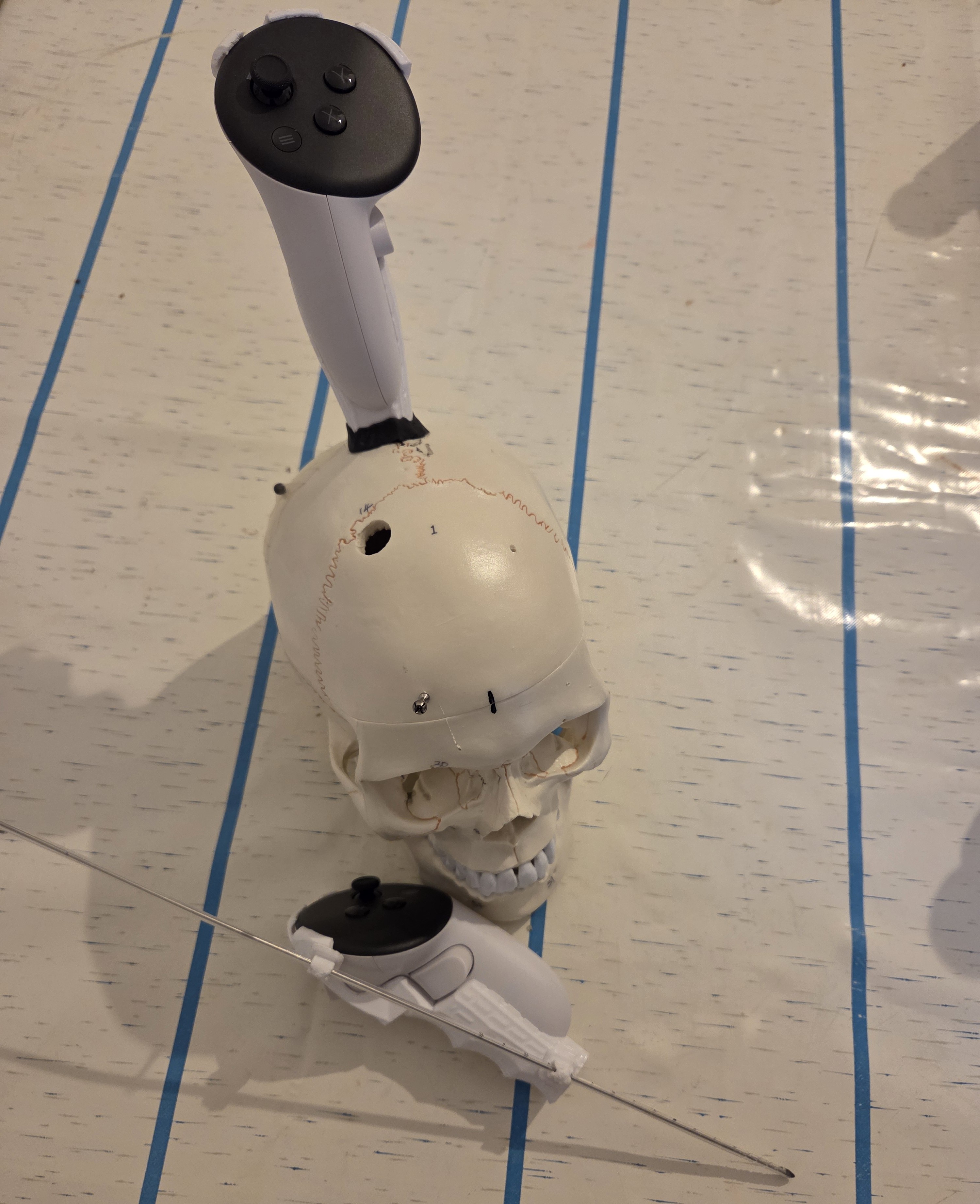}
        \caption{}
        \label{fig:Control1}
    \end{subfigure}
    \begin{subfigure}[b]{0.24\linewidth} 
        \centering
        \includegraphics[width=\linewidth]{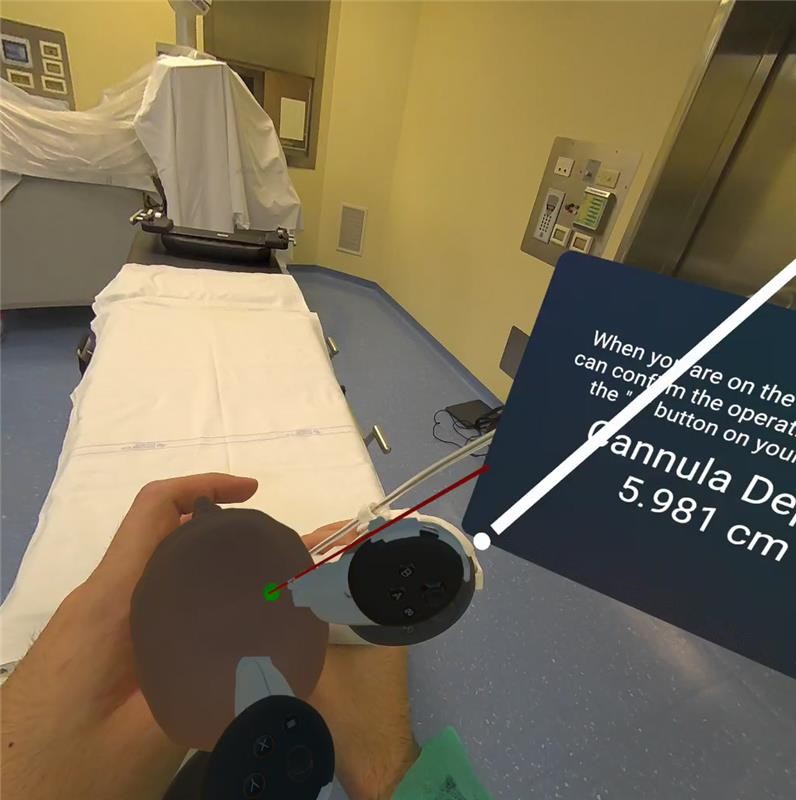}
        \caption{}
        \label{fig:Control2}
    \end{subfigure}
    \begin{subfigure}[b]{0.24\linewidth} 
        \centering
        \includegraphics[width=0.755\linewidth]{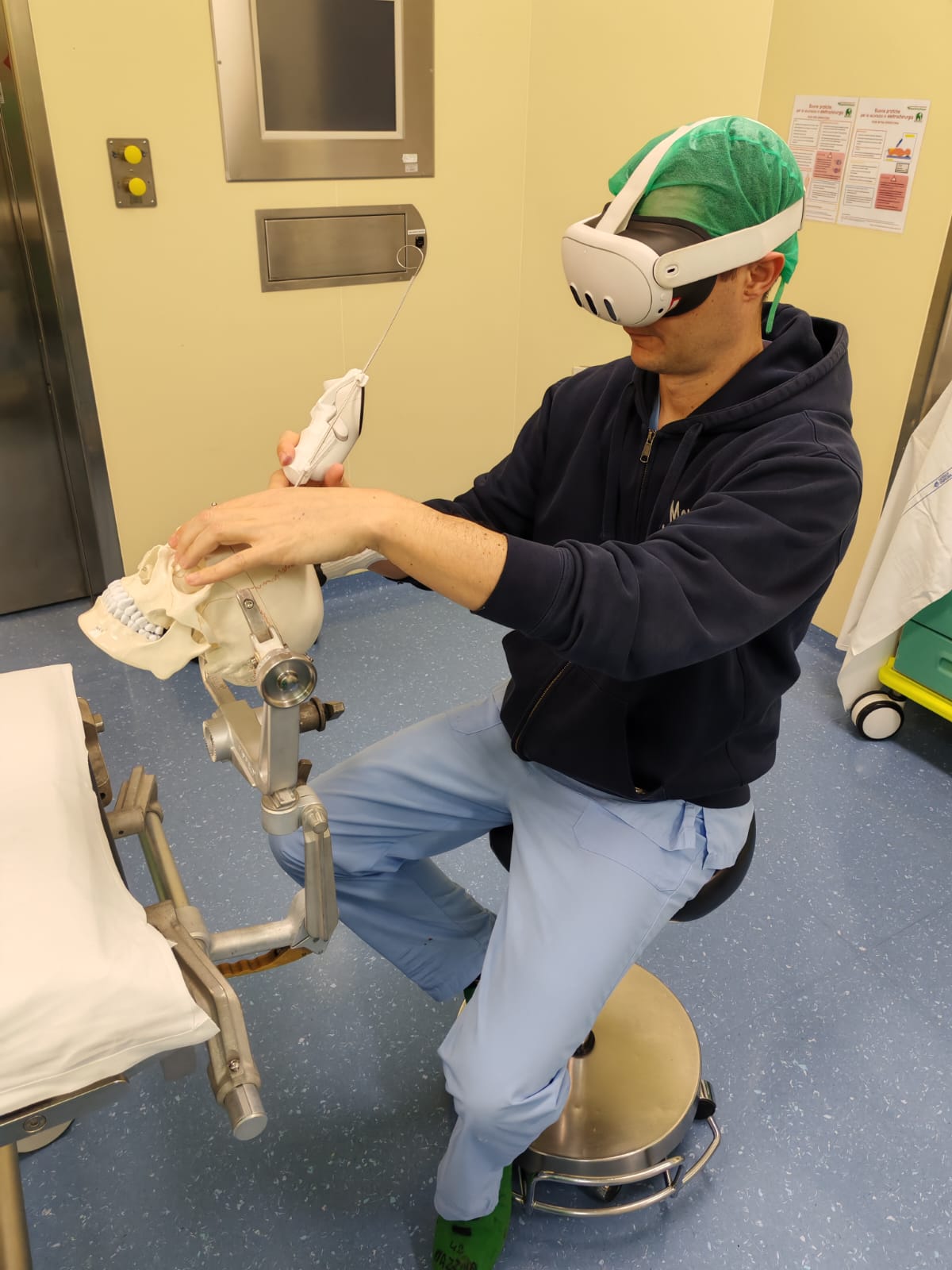}
        \caption{}
        \label{fig:Control4}
    \end{subfigure}
    \caption{A participant uses the testing system.}
    \label{fig:ControlAll} 
\end{figure}

Differently from the training system, the testing system considers a tracked Physical Catherer and a tracked Physical Skull (see \autoref{fig:Control1}). 
The Physical Catherer is a real graduated catheter used during EDV placement procedures. 
The markings on the Physical Catheter are intended to indicate the insertion depth to the user. 
The Physical Skull is a phantom skull. 
To provide haptic feedback during insertion, an agar-based material \cite{van_doormaal_development_2025} was placed inside the cranial cavity, simulating the resistance of brain tissue (gray matter). 
For the testing phase, the entry point (Kocher’s point) is predefined, thus removing the need for manual entry point selection. Hence the Physical Skull was perforated in advance.

A key technical challenge was achieving accurate tracking of the Physical Skull and the Physical Catheter due to restrictions imposed by Meta on direct camera access within the Meta Quest 3, thus preventing the usage of computer vision techniques as in \cite{robertson2021frameless}. 
However, camera-based tracking could have introduced latency (40–60 ms), GPU overhead (1–2\%), memory usage (45 MB per stream), a data rate capped at 30 Hz, and maximum resolution of 1280 $\times$ 960 pixels \footnote{see \href{https://developers.meta.com/horizon/documentation/unity/unity-pca-overview}{Meta’s documentation}}.
To overcome these limitations, we developed a custom tracking solution using Meta Quest 3 controllers, which provide real-time spatial tracking via inertial sensors and infrared LEDs. 
The controllers were mounted on the Physical Catheter and the Physical Skull with 3D-printed holders. This setup enabled accurate, privacy-preserving tracking at 60 Hz, outperforming Meta’s camera-based system in responsiveness and efficiency.

The 3D-printed mount was designed in other to let the user to held the Physical Catherer similarly to how a neurosurgeon would hold a catherer in a real scenario.
During the insertion, the user manipulates the tracked Physical Catheter directly using the hands, with the controller acting as a tracker. 
This allows for simultaneous control of insertion depth and angle, mimicking real surgical handling.  
By pressing the "A" button on the controller the user can confirm the desired target point is reached. 

The Application Engine runs within the Unity3D environment and is responsible for real-time visualization and synchronization of the Virtual Panel and the Digital Skull. Additionally, it handles the registration of the Digital Skull to its physical counterpart, the Physical Skull.
Concerning the Digital Skull, the same virtual head model used in the training system was rendered in the MR testing environment to provide consistent reference points established during the training phase. 
Since the graduated catheter is difficult to see clearly while wearing the headset, the insertion depth is also displayed in real-time on the Virtual Panel to support the user throughout the EVD placement (see \autoref{fig:Control2}).
Importantly, we remark that, in this setup, the MR interface was not used to provide any form of assistance (aids or feedback) to the user. This design choice ensures that the testing system can yield meaningful insights into each user's skill retention in an unaided scenario.
Differently from the training system, we point out that all performance measurements were computed in a virtual setting, while the procedere is performed with a real catherer on a real phantom skull (see \autoref{fig:Control4}). 
For example, the distance between the operation point (from the tip of the tracked Physical Catheter) and the virtual target point was calculated using the MR tracking system. 
This ensures high-fidelity evaluation of each trial while maintaining full control over data collection and eliminating manual measurement errors.

The Data Logger module collects and stores all interaction data during the EVD procedure.

\subsection{A Pilot Study for System Refinement}

A pilot study involving 10 participants was conducted to evaluate the initial version of the NeuroMix training and testing systems to identify potential issues in the experimental setup. 
None of these participants were included in the main experiment to prevent familiarity or learning-related biases. 
Among them, seven were neurosurgeons with varying levels of experience in EVD placement (one with low experience, two with moderate, and four with high experience) while the remaining three had no prior experience with neurosurgical procedures. 
All participants tested the No Aid, 2D Aid, and 2D-3D Aid modalities as well as the testing system. 
Their ages ranged from 24 to 45 years old. 
Concerning their familiarity with MR technologies, six had never used MR before and four reported moderate exposure.

During the pilot study sessions, some technical issues emerged, including application crashes due to a memory leak caused by rendering the high-resolution Digital Skull model, within both the training and testing systems. 
To address this, the 3D model was optimized in Blender to reduce complexity while preserving visual fidelity, and GPU-based rendering was implemented to offload image processing from the CPU, leveraging the Snapdragon XR2 Gen 2 architecture. 
These optimizations enabled the system to run at a full resolution of $2064 \times 2208$ pixels per eye with a stable 90 Hz refresh rate, ensuring smooth and responsive performance.

During the testing phase, although the tracked Physical Catheter featured graduated markings, participants had difficulty seeing them clearly in passthrough mode \cite{guo2022augmented}. 
To address this, the insertion depth values were displayed in real-time on the Virtual Panel.
Additionally, due to camera distortion in passthrough mode \cite{guo2022augmented}, users experienced some difficulties when inserting the Physical Catheter into the hole at the Kocher’s point on the Physical Skull. 
To mitigate this issue, a digital red stick was registered to the Physical Catheter to provide a clearer visual reference.

In terms of usability, users were initially required to interact with virtual objects in the training system using controllers, which many found to be cumbersome.
Additionally, the only method for selecting the Rotation and Tilt components of the Digital Catheter was through sliders on the Virtual Panel.
To improve ease of interaction, hand tracking was introduced, allowing users to directly manipulate the Virtual Catheter with their hands.
These adjustments align with findings in the literature \cite{jeffri2021review} emphasizing that reducing cognitive load through effective interface and interaction modality design \cite{mendes2019survey} is essential to ensure both improved user performance and reliable experimental results, particularly in tasks requiring high levels of spatial reasoning.

\subsection{Study design}

This section outlines the experimental design, procedure, participants, and measurements used in the study.

\subsubsection{Study Objective}

The experiment employed a between-subjects design to assess the effectiveness of the developed training and testing systems, as well as the impact of three different training modalities (No Aid, 2D Aid, and 2D-3D Aid) offering varying levels of visual aids on short-term skill retention for the EVD placement procedure.
Participants were divided into two main groups: the Experimental Group and the Control Group. The Experimental Group completed both the training and testing phases, whereas the Control Group took part only to the testing phase.
Specifically, the experiment aims to assess how different learning modalities influence users’ perceived usability, cognitive workload, and acceptance of the training system during the training phase. Additionally, the study investigates the impact of each modality on procedural precision and execution time, both during training and in a subsequent testing phase in which visual aids are no longer available. A comparison with a Control Group, which did not receive training through the developed system, is included to evaluate the retention effects more rigorously.

\subsubsection{Procedure and Task}
We now outline the experimental procedure followed by the Experimental Group and the Control Group (see \autoref{fig:System}, right panel). 
Participants were randomly assigned to one of the two groups. 
All participants were asked to complete a demographic questionnaire, which included information such as age, gender, and education level.
This was followed by a background questionnaire aimed at assessing their prior experience with EVD procedures, neuronavigation systems, interpretation of brain CT DICOM images, the use of software such as 3D Slicer, and their familiarity with various immersive technologies. 

Then, participants in the Experimental Group were randomly assigned to one of the three learning modalities: No Aid, 2D Aid, or 2D-3D Aid. 
For each participant in the Experimental Group, the experiment consisted of three distinct phases: the Familiarization Phase, the Training Phase, and the Testing Phase.

During the Familiarization Phase all participants took part in an introductory session led by an experienced operator, who explained the theoretical and practical aspects of the EVD procedure according to standard neurosurgical protocols \cite{flint2013simple}.
This session covered topics such as the clinical indications for EVD placement, common anatomical landmarks (with an emphasis on Kocher’s point and Monro's foramen), the ideal catheter trajectory and depth (generally 5 to 7 cm), confirmation cues for correct catheter positioning (e.g., release of cerebrospinal fluid), and potential complications including infections, hemorrhages, device failure, and misplacement.
Following this, participants were guided through a familiarization session specific to the learning modality they had been assigned to. During this session, they were introduced to the system’s features and instructed on how to interact with the Virtual Panel, Digital Catherer, and Digital Skull provided by their respective setup.

After the Familiarization Phase, the Training Phase begins. 
During this phase the participants interact exclusively with virtual tools \cite{seymour2002virtual}. 
Each participant is asked to perform five virtual EVD placements. Depending on the assigned learning modality, participants practice inserting the Digital Catherer into the Digital Skull towards the the target point (Monro's foramen) highlighted  on the CT scans. 
Depending on the training modalities, the procedure is supported by visual aids, when available, in order to assist partecipants in selecting the correct entry point and achieving proper catheter placement. After each attempt, participants receive feedback tailored to their assigned modality. 
This feedback is designed to help them understand how to improve their technique, in line with evidence from the literature supporting the role of feedback in enhancing procedural learning \cite{eom2024did, butaslac2022systematic}.

Once the Training Phase is completed, the Testing Phase begins. 
In this phase, participants interact with the Physical Skull, a phantom model enhanced for realism by filling the cranial cavity with an agar-based material that mimics the consistency of brain tissue. This setup provides users with haptic feedback, simulating the EVD placement procedure in a more lifelike manner. 
The entry point is indicated to the participant by a pre-existing borehole on the Digital Skull.
To ensure visual continuity and consistency, the system registers the same Digital Skull used during the Training Phase on the Physical Skull.
A Physical Catheter, identical to the one used in real surgical procedures, is also employed.
This allows participants to perform the EVD procedure using physical tools on the same clinical data they trained with virtually. As a result, the anatomical landmarks remain unchanged from the training phase \cite{greenberg_anatomy_2023, flint2013simple}.
Participants are required to perform five EVD placements during this phase. No visual aids or feedback are provided during this phase.

The Control Group did not undergo any virtual Training Phase but performed five EVD placements as described for the Testing Phase performed by the Experimental Group. 
However, prior to the insertions, the Control Group was allowed to visualize the CT scans on a standard computer monitor using Slicer3D for reference.

At the end of the experiment, all the participants were asked to complete questionnaires depending on their group. 

\subsubsection{Participants}

A total of 48 participants took part in the main experiment, ranging in age from 23 to 38 years (M = 28.04, SD = 3.13), including 21 females and 28 males. 
This sample size aligns with prior research in the field \cite{buwaider_augmented_2024,butaslac2022systematic}.
Results from the background questionnaire indicated that all participants had a medical background and were selected among recently hired clinicians with limited experience in EVD placement (M = 2.00, SD = 0), limited familiarity with neuronavigation systems (M = 1.19, SD = 0.57), and moderate experience in interpreting neuroimaging such as CT and MRI scans (M = 3.41, SD = 0.69).
In terms of handedness, 43 participants were right-handed, 4 left-handed, and 1 ambidextrous.
Regarding familiarity with immersive technologies, participants had limited familiarity with Augmented, Mixed, or Virtual Reality (M = 2.57, SD = 0.68). Specifically, the vast majority (45 out of 48) had no prior experience using the Meta Quest 3 device employed in this study, and limited experience with other head-mounted displays (M = 2.60, SD = 0.63).
Of the 48 participants, 36 were assigned to the Experimental Group and evenly distributed across the three training modalities: No Aid (n = 12), 2D Aid (n = 12), and 2D-3D Aid (n = 12). The remaining 12 participants were assigned to the Control Group.

\subsubsection{Measurements}

Various objective and subjective metrics were collected to assess the impact of the NeuroMix training on procedural accuracy and efficiency. 
Concerning objective measures, during the Training Phase the system recorded: (a) the number of attempts required to identify the correct entry point, (b) the 3D coordinates of the operation points, (c) the Rotation and Tilt components selected, and (d) the Total Time (TT) for each insertion.
During the Testing Phase, only metrics (b), (c) and (d) were recorded, as the entry point was predefined and provided to participants.
From (b) we derived the Error Score (ES) computing the Euclidean distance between the operation point and the target point. From (c) we derived the Rotation Error (RE) and the Tilt Error (TE) which are defined as the absolute angular differences between the user-selected Rotation and Tilt and the ideal trajectory toward the target point. 
For each participant and training modality, we computed the Average metric by averaging the individual error values (e.g., Error Score, Tilt Error, Rotation Error, Total Time) across all trials (5 trials for the Training Phase and 5 trials for the Testing Phase for each participant).

Subjective measures were measured through standardized questionnaires.  
The Experimental Group completed the System Usability Scale (SUS), a 10-item questionnaire rated on a 5-point Likert scale. The items are grouped into two dimensions: Usability and Learnability \cite{lewis2009factor}. 
In addition to analyzing these two subscales, we also calculated the Total score as the sum of all item responses across the entire SUS questionnaire.
In addition, the Experimental Group completed the NASA Task Load Index (NASA-TLX) \cite{HART1988139}, which is based on a 10-point Likert scale and evaluates six dimensions of perceived workload: Mental Demand, Physical Demand, Temporal Demand, Performance, Effort, and Frustration. We also calculated the Average workload score by averaging across all dimensions.
The Experimental Group was asked to complete the Technology Acceptance Model (TAM) questionnaire \cite{davis1989perceived} was also administered, using a 5-point Likert scale. It includes two core dimensions: Perceived Usefulness and Perceived Ease of Use. 
Finally, for the Experimental Group we developed a custom Final Test Assessment (FTA) questionnaire to evaluate whether participants' perceived complexity of the Testing Phase aligned with their actual performance. 
The questionnaire consisted of five items, each rated on a 5-point Likert scale (e.g., "How do you rate the difficulty of accomplishing the X insertion of the catheter during the testing phase?").

The Control Group completed only the FTA questionnaire as they did not undergo the virtual training.

\subsubsection{Tools for Statistical Analysis}

For the statistical analysis, we primarily relied on a set of Python libraries designed for data manipulation, statistical modeling, and visualization. These included Pandas (v2.2.3), Numpy (v2.2.3), Scipy (v1.15.2), Statsmodels (v0.14.4), Patsy (v1.0.1), Matplotlib (v3.10.1), Seaborn (v0.13.2) and Plotly (v6.0.1).
The goal of the analysis was to identify significant differences or similarities both within the Experimental Group (across the three training modalities) and between the Experimental and Control Groups. We followed the statistical protocol outlined below. 
First, the normality of the data distributions was assessed using the Shapiro-Wilk test \cite{razali2011power}. If normality was confirmed, a one-way ANOVA was applied to detect significant differences between groups, followed by pairwise comparisons using independent samples t-tests \cite{lakens2013calculating}. 
To evaluate statistical equivalence, we performed a two-sided TOST (Two One-Sided Tests) procedure \cite{lakens2017equivalence}. 
If normality was not satisfied, non-parametric tests were used: the Kruskal-Wallis H-test \cite{mckight2010kruskal} to assess differences across groups and the Mann-Whitney U test for pairwise comparisons \cite{mcknight2010mann}. Finally, we employed the Analysis of Covariance (ANCOVA) \cite{van2013ancova} test to examine the effect of training modality on procedural performances while controlling for potential covariates.

\section{Results}

This section reports the outcomes of the statistical analyses, distinguishing between subjective and objective metrics.

\subsection{Subjective Measures}

We here present the results of the subjective evaluations collected through the standardized post-experience questionnaires.

\subsubsection{SUS questionnaire}

In \autoref{table:meanssd} we report the mean (M) and standard deviation (SD) by construct of the SUS questionnaire for each of the learning modalities.
All the modalities were given high Total values ($>$70).
Normality was confirmed for each of the constructs across all the learning modalities.
The ANOVA test did not show any significant differences among the three modalities for Usability (p=0.521), Learnability (p=0.428) and Overall (p=0.499). 
A pair-wise TOST analysis revealed significant equivalence in terms of Usability: No Aid vs. 2D Aid (p = 0.012), 2D Aid vs. 2D-3D Aid (p = 0.035), and No Aid vs. 2D-3D Aid (p = 0.046).

\subsubsection{NASA questionnaire} 

The mean and standard deviation of each NASA dimension are reported in \autoref{table:meanssd}.
Normality was achieved by the Physical, Temporal, and Effort dimensions, with the Mental, Performance, and Frustration dimensions failing to meet the normality assumption.
A one-way ANOVA did not reveal significant differences among the three learning modes in the Physical, Temporal, and Effort dimensions.
However, the Kruskal–Wallis H test revealed a significant difference in Frustration scores (p = 0.049), with no differences found for the Mental and Performance dimensions.
Pair-wise comparisons using the Mann–Whitney U test revealed a significant difference for No Aid vs 2D Aid (p = 0.038), but not for No Aid vs 2D-3D Aid (p = 0.081), or for 2D Aid vs 2D-3D Aid (p = 0.498).
Finally, the TOST procedure did not reveal any significant equivalences between the modalities. 

\subsubsection{TAM questionnaire}

Detailed scores for the TAM metric, including means and standard deviations for all dimensions, are reported in \autoref{table:meanssd}. 
For all these dimensions of the TAM questionnaire, the normality was achieved. 
The ANOVA test did not highlight significant differences for Usefulness (p=0.370) and for Ease of Use (p=0.183) among the three learning modalities.
A pair-wise TOST test showed significant equivalence for the Ease of Use construct for No Aid vs 2D Aid (p=0.034), No Aid vs 2D-3D Aid (p=0.046) and 2D Aid vs 2D-3D Aid (p=0.0001). Moreover, significant equivalence was found when applying the TOST test to No Aid vs 2D Aid (p=0.021) for the Usefulness construct.

\subsubsection{FTA questionnaire}

In \autoref{table:meanssd}, we reported the mean and standard deviation scores for the Overall FTA questionnaire. A Kruskal–Wallis H test conducted across all groups did not reveal any statistically significant differences. However, the TOST analysis indicated a statistically significant equivalence between the No Aid and 2D Aid groups, while no equivalence was found between the other group pairs.


\begin{table}[h!]
\centering
\caption{Summary of Subjective Measures.}
\renewcommand{\arraystretch}{0.9}
\scalebox{1}{
\begin{tabular}{|l|l|c|c|c|c|}
\hline
\textbf{Variable} & \textbf{Construct} & \textbf{No Aid} & \textbf{2D Aid} & \textbf{2D-3D Aid} & \textbf{Control} \\
& & \textbf{(M, SD)} & \textbf{(M, SD)} & \textbf{(M, SD)} & \textbf{(M, SD)} \\
\hline
\multirow{3}{*}{\textbf{SUS}} & Usability     & 4.24, 0.47 & 4.16, 0.37 & 4.02, 0.55 & -- \\
                              & Learnability  & 3.88, 0.77 & 3.88, 0.68 & 3.54, 0.69 & -- \\
                              & Total   & 91.67, 11.55 & 90.00, 10.11 & 85.62, 13.86 & -- \\
\hline
\multirow{7}{*}{\textbf{NASA-TLX}} & Mental       & 5.42, 1.88 & 6.17, 1.34 & 6.25, 1.48 & -- \\
                                   & Physical     & 3.17, 1.70 & 3.25, 1.71 & 4.25, 1.71 & -- \\
                                   & Temporal     & 3.75, 2.05 & 4.00, 1.28 & 3.67, 1.44 & -- \\
                                   & Performance  & 7.33, 1.61 & 6.67, 1.92 & 7.58, 0.79 & -- \\
                                   & Effort       & 4.92, 2.23 & 6.17, 1.85 & 6.08, 1.83 & -- \\
                                   & Frustration  & 2.00, 1.13 & 3.58, 2.07 & 3.17, 2.08 & -- \\
                                   & Average & 4.43, 0.89 & 4.97, 0.84 & 5.17, 1.02 & -- \\
\hline
\multirow{2}{*}{\textbf{TAM}} & Usefulness    & 4.18, 0.62 & 4.17, 0.48 & 4.45, 0.51 & -- \\
                              & Ease of Use   & 2.85, 0.41 & 3.07, 0.30 & 3.09, 0.32 & -- \\
\hline
\multirow{1}{*}{\textbf{FTA}} & Average & 2.98, 0.78 & 2.95, 0.36 & 3.07, 0.83 & 2.75, 0.86 \\
\hline
\end{tabular}
}
\label{table:meanssd}
\end{table}

\subsection{Objective measures}

This section presents the statistical analysis results of the objective metrics collected during the Training phase and the Testing phase.

\subsubsection{Accuracy analysis: Entry, Target and Operation Points}

Regarding Entry Point selection, in $94\%$ of the total instances, the correct option (Kocher’s point) was identified on the first attempt during the Training Phase.

In \autoref{table:combined_operation_precision} we report the mean and standard deviation of the Average Error Scores for the training and testing phases, along with the differences in percentage between Training and Testing (red), and the differences percentage between each learning modality and the Control group (blue).

The results of the Shapiro-Wilk test indicated that all the variables violated the assumption of normality.
The Kruskal–Wallis H-test revealed statistically significant differences between the three learning modalities during the Training phase (p = 0.0018).
The Mann–Whitney U test showed the following significant differences: No Aid vs 2D Aid (p = 0.0001), No Aid vs 2D-3D Aid (p = 0.0001), and 2D Aid vs 2D-3D Aid (p = 0.011), suggesting improvements in precision with increased system support.

For what concerns the Testing Phase the four modalities (including Control) showed significant difference (p =0.0023), whereas pairwise comparisons showed that the 2D-3D Aid condition differed significantly from all others: No Aid (p = 0.0053), 2D Aid (p = 0.00021), and Control (p = 0.00014). 
Significant differences were observed between the Control and the No Aid groups (p = 0.001). 
No significant differences were found between 2D Aid and No Aid (p = 0.184), and between Control e 2D Aid (p = 0.068).

\begin{figure*}[!htb]
    \centering
    \begin{subfigure}[b]{0.3\linewidth}  
        \centering
        \includegraphics[width=0.985\linewidth]{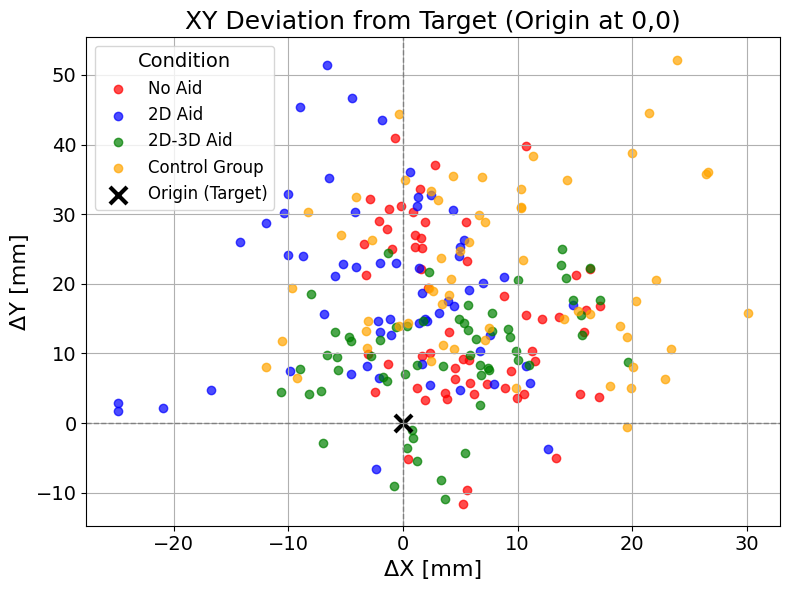}
        \caption{}
        \label{fig:scatter1}
    \end{subfigure}
    \begin{subfigure}[b]{0.3\linewidth} 
        \centering
        \includegraphics[width=0.985\linewidth]{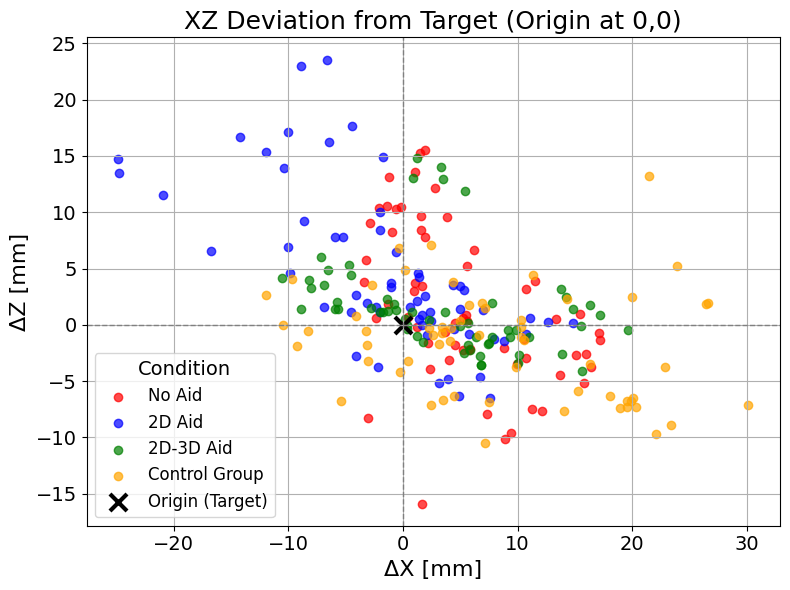}
        \caption{}
        \label{fig:scatter2}
    \end{subfigure}
    \begin{subfigure}[b]{0.3\linewidth} 
        \centering
        \includegraphics[width=0.985\linewidth]{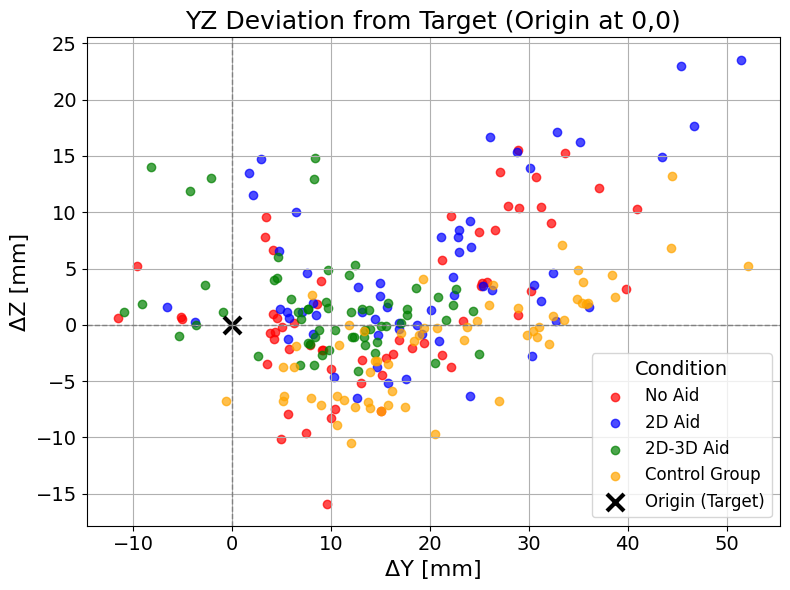}
        \caption{}
        \label{fig:scatter3}
    \end{subfigure}
    \caption{Spatial deviations from the target across sagittal [XY], coronal [XZ], and axial [YZ] planes.}
    \label{fig:scatterplot}
\end{figure*}

To perform a more detailed analysis on the Testing Phase data, we computed the component-wise absolute differences between the target point and the operation point along the sagittal X, axial Y, and coronal Z axes. 
In the sagittal direction X, significant differences were observed only between the Control condition and the other three modalities: No Aid (p = 0.0012), 2D Aid (p = 0.0040), and 2D-3D Aid (p = 0.0188).
In the axial direction Y, more widespread differences emerged, particularly involving the 2D-3D Aid condition, which significantly differed from all others: 2D Aid (p = 0.00012), Control (p = 5.02e-7), and No Aid (p = 0.036). 
Similarly, in the coronal direction Z, the 2D-3D Aid modality showed significantly differences when compared to No Aid (p = 0.00066), 2D Aid (p = 0.00399), and Control (p = 0.0181).
For the Testing Phase, \autoref{fig:scatterplot} illustrates the spatial deviations from the target point (fixed in the origin) and the operation points across the three main anatomical planes: sagittal [XY], coronal [XZ], and axial [YZ].
In all the views, the 2D-3D Aid condition (green) shows the highest concentration of data points around the target. 
In contrast, the Control condition (orange) displays the widest dispersion, particularly along the positive vertical Y axis (see \autoref{fig:scatter1}, \autoref{fig:scatter3}). 
The No Aid (red) and 2D Aid (blue) conditions show intermediate behavior. 
From visual inspection, Control, No Aid and 2D Aid groups seem to show a deviation along the positive axial direction Y, differently from the 2D-3D Aid group. 

Finally, we conducted an ANCOVA test using the number of training attempts as a covariate, the training modality as a between-subject factor and precision during Testing Phase as dependent variable. 
The analysis revealed a significant main effect of training modalities on operation precision (p= 0.0021), indicating that the type of training used had an impact on precision, independent of the number of attempts.

A Spearman correlation analysis \cite{de2016comparing} was conducted to investigate the relationship between the FTA metric (\autoref{table:meanssd}) and the operation precision (\autoref{table:combined_operation_precision}) during the Testing Phase. The results revealed a weak but statistically significant (p = 0.012) negative correlation, indicating that higher levels of perceived complexity were associated with lower precision scores.


\begin{table}[h!]
\centering
\caption{Summary of the Average Error Scores [cm].}
\scalebox{1}{%
\begin{tabular}{|c|l|c|c|c|c|}
\hline
\textbf{Phase} & \textbf{Metric} & \textbf{No Aid} & \textbf{2D Aid} & \textbf{2D-3D Aid} & \textbf{Control} \\
 &  & \textbf{(M, SD)} & \textbf{(M, SD)} & \textbf{(M, SD)} & \textbf{(M, SD)} \\
\hline
\multirow{1}{*}{\textbf{Training}} 
 & Average ES & 1.05, 0.57 & 0.67, 0.65 & 0.42, 0.27 & -- \\
\hline
\multirow{1}{*}{\textbf{Testing}} 
 & Average ES & 2.00, 0.98 & 2.30, 1.15 & 1.47, 0.60 & 2.64, 1.06\\
\hline
\hline
 &  Train / Test (\%) & \textcolor{red}{+90.5\%} & \textcolor{red}{+243.28\%} & \textcolor{red}{+259\%} & -- \\
 & Mod. / Control (\%)  & \textcolor{blue}{-24.2\%} & \textcolor{blue}{-12.9\%} & \textcolor{blue}{-44.3\%} & -- \\
\hline
\end{tabular}
}
\label{table:combined_operation_precision}
\end{table}

\begin{table}[h!]
\centering
\caption{Summary of Avarage Tilt and Rotation Errors [$^\circ$]}
\scalebox{1}{%
\begin{tabular}{|c|l|c|c|c|c|}
\hline
\textbf{Phase} & \textbf{Metric} & \textbf{No Aid} & \textbf{2D Aid} & \textbf{2D--3D Aid} & \textbf{Control} \\
 &  & \textbf{(M, SD)} & \textbf{(M, SD)} & \textbf{(M, SD)} & \textbf{(M, SD)} \\
\hline
\multirow{2}{*}{\textbf{Training}} 
 & Average TE     & 4.94, 3.87 & 4.08, 5.33 & 1.74, 1.58 & -- \\
 & Average RE  & 5.05, 3.92 & 2.78, 2.21 & 2.20, 1.86 & -- \\
\hline
\multirow{2}{*}{\textbf{Testing}}  
 & Average TE    & 14.60, 9.97 & 16.94, 10.58 & 10.04, 5.08 & 17.97, 10.12 \\
 & Average RE & 6.02, 4.40  & 7.27, 6.41   & 6.14, 4.21  & 9.44, 7.25 \\
\hline
\hline                  
\multirow{2}{*}{} 
 & Train / Test (Tilt)     & \textcolor{red}{+195.34\%} & \textcolor{red}{+315.20\%} & \textcolor{red}{+477.01\%} & -- \\
 & Train / Test (Rotation) & \textcolor{red}{+19.21\%}  & \textcolor{red}{+161.51\%} & \textcolor{red}{+179.09\%} & -- \\
\hline
\multirow{2}{*}{} 
 & Mod. / Control (Tilt)     & \textcolor{blue}{-18.76\%} & \textcolor{blue}{-5.73\%}  & \textcolor{blue}{-44.10\%} & -- \\
 & Mod. / Control (Rotation) & \textcolor{blue}{-36.22\%} & \textcolor{blue}{-22.96\%} & \textcolor{blue}{-34.96\%} & -- \\
\hline
\end{tabular}
}
\label{table:combined_angle_diff}
\end{table}

\subsubsection{Accuracy analysis: Tilt and Rotation Components}

The mean and standard deviation values for the Average Tilt Error and for the Average Rotation Error are reported in \autoref{table:combined_angle_diff}.  
The red values indicate the difference in percentage between Training and Testing, whereas the blue values represent the difference in percentage between each learning modality and the Control group.

A Kruskal–Wallis H test performed on the Training Phase data revealed significant differences in angular accuracy among the three learning modalities, both for Tilt (p = 0.0034) and Rotation (p = 0.012).
Pairwise comparisons using the Mann–Whitney U test confirmed significant differences between all pairs of modalities.

The Kruskal–Wallis H test performed on the Testing Phase data revealed significant differences among the No Aid, 2D Aid, 2D-3D Aid and Control groups for the Tilt component (p=0.0042). No significant differences were found for the Rotation component (p=0.0092).  
For the Tilt-related data, the Mann–Whitney U test confirmed that the 2D–3D Aid condition was significantly different if compared to the 2D Aid (p=0.007) and the Control (p= 0.005), while it was not significant different from the No Aid (p = 0.071). Additionally, the No Aid condition was significantly different from Control (p = 0.036), whereas it did not show significative difference with respect to 2D Aid (p=0.067).

Finally, we conducted an ANCOVA using the number of training attempts as a covariate, the training modality as a between-subject factor, and Rotation and Tilt Errors during the Testing Phase as the dependent variable. 
For the Rotation errors, the analysis revealed no significant effect of the training modalities (p = 0.338) while the number of attempts had a significant impact (p=0.0004).
In contrast, for the Tilt component, both the training modalities (p = 0.0056) and the number of attempts (p=0.0002) had a significant main effect, indicating that the type of training aid influenced tilt accuracy beyond mere repetition.

\subsubsection{Time Analysis}

We report in \autoref{table:combined_timeoperation} the mean and standard deviation values of the Average Total Time. 
The percentage variations are also highlighted: red values indicate difference in percentage between training to testing, while blue values indicate difference in percentage between each of the learning modality and the Control group.

A Kruskal-Wallis H test was conducted to compare the Average Total Time on the training phase across the three learning modes (No Aid, 2D Aid, and 2D-3D Aid). 
The test was statistically significant (p= 0.0001), indicating that the total time required to complete the operation differed across modes. 

A Kruskal-Wallis H test was conducted to evaluate differences of the Average Total Time for the Testing Phase across No Aid, 2D Aid, 2D-3D Aid and Control groups. 
The analysis revealed a statistically significant difference within the data ($p = 0.003$), suggesting that the type of aid used had a measurable effect on the operation time.

Finally, we conducted an ANCOVA using the number of training attempts as a covariate, the training modality as a between-subject factor, and the Total Time during the Testing Phase as the dependent variable. 
The training modality did not show a significant main effect (p = 0.396) on operation time while the number of training attempts did (p = 0.0045). 

\begin{table}[h!]
\centering
\caption{Summary of the Average Total Time [s].}
\scalebox{1}{
\begin{tabular}{|c|l|c|c|c|c|}
\hline
\textbf{Phase} & \textbf{Metric} & \textbf{No Aid} & \textbf{2D Aid} & \textbf{2D-3D Aid} & \textbf{Control} \\
 &  & \textbf{(M, SD)} & \textbf{(M, SD)} & \textbf{(M, SD)} & \textbf{(M, SD)} \\
\hline
\multirow{1}{*}{\textbf{Training}} 
 & Average TT & 62.21, 46.36 & 137.84, 78.90 & 132.09, 93.25 & -- \\
\hline
\multirow{1}{*}{\textbf{Testing}} 
 & Average TT & 42.87, 23.36 & 47.25, 20.64 & 49.74, 37.04 & 33.63, 14.53 \\
\hline
\hline
 & Train / Test (\%) & \textcolor{red}{-31.08\%} & \textcolor{red}{-65.70\%} & \textcolor{red}{-62.40\%} & -- \\
 & Mod. / Control (\%) & \textcolor{blue}{+27.44\%} & \textcolor{blue}{+40.48\%} & \textcolor{blue}{+47.88\%} & -- \\
\hline
\end{tabular}
} 
\label{table:combined_timeoperation}
\end{table}

\section{Discussion}
In this section, we seek answers to the RQs.\\ \textit{RQ1: How do users' perceived usability (measured by SUS), workload (measured by NASA-TLX), and technology acceptance (measured by TAM) differ when comparing the No Aid, 2D Aid, and 2D-3D Aid groups in EVD placement during the training phase?}.

All the three training modes (No Aid, 2D Aid, and 2D–3D Aid) were found to be highly acceptable and usable to the participants.
SUS scores were always more than the usual acceptance level for usability of 70 \cite{bangor2009determining} without any significant differences between the groups and showing significant equivalence within all the paired comparisons. 
These findings are presented in \autoref{table:meanssd} and show that even highly advanced visualizations, such as animated 3D trajectory guidance, do not negatively impact perceived usability. Similarly, the TAM measures revealed high scores for Perceived Usefulness and Perceived Ease of Use for all modalities (see \autoref{table:meanssd}), and statistical equivalence between all pairs was determined based on Ease of Use dimension.
This is consistent with prior research showing that immersive training systems can maintain high usability when interface design is intuitive and familiar \cite{linte2013mixed,mendes2019survey,wells2024technical}. 
The interface design was inspired by clinical neuronavigation systems \cite{gumprecht1999brainlab,alazri2017placement}, which primarily rely on CT imaging that clinicians are typically familiar with due to their experience.

Regarding workload, measured through the NASA-TLX, most dimensions showed no significant differences. However, the Frustration dimension was significantly higher for the 2D Aid condition compared to No Aid.
This may reflect the higher cognitive effort required to interpret 2D projections without spatial context. In contrast, the 2D–3D Aid group, despite receiving more information, did not show increased frustration or significantly higher workload, suggesting that well-integrated 3D aids may help reduce cognitive effort through intuitive spatial cues \cite{lin2021labeling,wenk2023effect}. 

\textit{RQ2: How do users' procedure precision and execution time differ when comparing the No Aid, 2D Aid, and 2D-3D Aid groups in EVD placement during the training phase?} During the training phase, both catheter placement precision and execution time varied significantly depending on the visual support modality (see \autoref{table:combined_operation_precision},\autoref{table:combined_angle_diff}, \autoref{table:combined_timeoperation}).
Participants trained with 2D and 2D–3D visual aids achieved significant higher accuracy compared to the No Aid group (see \autoref{table:combined_operation_precision}). The highest accuracy was reached by the 2D-3D Aid group. The ANCOVA test controlling for the number of attempts indicated that the improvements were attributable to the training modalities, rather than practice effects. 
We point out that although the 2D Aid group achieved low error rates during training, it reported significant high levels of Frustration on the NASA-TLX scale unlike the 2D–3D Aid group.
These results are consistent with prior studies demonstrating that visual overlays, particularly when combining 2D imaging with 3D spatial cues, can improve performance in image-guided neurosurgical procedures \cite{alizadeh_virtual_2024,buwaider_augmented_2024}. 
Angular accuracy also improved, with the 2D–3D Aid group participants having significantly lower Tilt and Rotation errors, suggesting that the 3D trajectory guidance facilitated mental rotation and alignment, a benefit of immersive training systems for procedural tasks \cite{butaslac2022systematic, linte2013mixed}.
However, the execution time more than doubled in the aided groups. This reflects the well-known trade-off between performance accuracy and task efficiency observed in high-fidelity simulations and not only for procedural tasks \cite{buwaider_augmented_2024,daling2024effects}. 

\textit{RQ3: What is the impact of the No Aid, 2D Aid, and 2D-3D learning modalities for EVD placement on users' skill retention, measured by procedure accuracy and execution time, once these aids are no longer provided?} 
The testing phase was specifically designed to evaluate short-term skill retention by removing all MR-based visual aids and requiring participants to perform EVD placement using only physical instruments. 
The results show that prior exposure to training, particularly when enhanced with combined 2D and 3D guidance, led to a substantial improvement in unaided performance (see \autoref{table:combined_operation_precision}).  
Participants trained with 2D–3D Aid retained the lowest error in terms of Euclinead Distance during the Testing Phase, outperforming also the Control group. 
A significant difference was observed along the axial (Y) axis between the 2D–3D Aid group and the No Aid, 2D Aid, and Control groups.
Interestingly, from visual inspection we observe a noticeable shift towards positive Y values in the No Aid, 2D Aid, and especially the Control groups (see \autoref{fig:scatterplot}). In contrast, the 2D–3D Aid group demonstrated more precise performance, with operation points more evenly distributed around the target.
These results indicate that the integration of 2D and 3D spatial guidance during training enhances procedural memory, allowing users to mentally reconstruct accurate trajectories even without external aids. 
This aligns with existing research highlighting the role of immersive spatial encoding in supporting transfer of motor skills \cite{butaslac2022systematic,daling2024effects, lin2021labeling}. 
Moreover, an ANCOVA controlling for the number of training attempts confirmed a significant effect of training modality on retention performance, emphasizing that the quality, not just the quantity, of training matters \cite{sels2002more}. 
Interestingly, although the 2D–3D Aid group had the best accuracy, it did not translate to faster execution. 
All trained groups were slower than the Control group during unaided testing. This may reflect a more cautious, deliberate execution strategy, commonly observed after high-fidelity training, where accuracy is prioritized over speed \cite{batmaz2016getting}. 
Finally, the ANCOVA test suggested that while different training modalities significantly impacted skill retention in terms of accuracy, especially when using advanced combined 2D and 3D visual aids, speed rely more on repeated practice. 
We assume this is due to the fact that participants were not encouraged to prioritize speed during the training phase, but rather to focus on accuracy. Indeed, all the feedback provided during training related solely to placement precision, with no emphasis on execution time. Even when using the 2D–3D Aid, which offered 3D visualization of the trajectory, participants still required a relatively long time on average \cite{vekony2022speed}.

\section{Limitations and Future Works}

This study demonstrated the effectiveness of the NeuroMix system for EVD placement training, however we acknowledge some limitations.
The evaluation focused exclusively on short-term skill retention, as the Testing Phase was conducted immediately following the Training Phase. Future work will aim to investigate long-term retention effects to better understand how procedural skills evolve over time.
Although the Testing Phase involved a physical catheter and skull phantom, it lacked important aspects of clinical realism, such as dynamic tissue deformation, bleeding, and the pressures of time-sensitive decision-making. 
These factors are known to influence user behavior, stress levels, and performance in real-world settings \cite{zhou2024time}. 
Moreover, the Training Phase was conducted on a single digital skull model, which may have limited the expressiveness of the system by not challenging users with the anatomical variability commonly encountered in clinical practice. Similarly, the use of a single patient-specific CT dataset restricted the diversity of training scenarios, preventing participants from experiencing a broader range of anatomical and pathological conditions.
To address these issues, future developments will include the integration of multiple clinical cases and CT datasets, along with the 3D printing of anatomical models to enhance the realism of both the training and testing phases and to assess the impact on skill retention.
Additionally, the tracking system used in this study was based on controller-mounted sensors, which while being practical and efficient may not reach the precision levels of optical navigation systems used in operating rooms. As Meta has recently enabled access to its headset cameras, we plan to upgrade the system to incorporate advanced computer vision algorithms for markerless tracking.

\section{Conclusions}

This study introduced NeuroMix, a Mixed Reality simulator for EVD placement, designed to assess how different visual aids affect procedural performance and short-term skill retention.
In a controlled study with 48 participants, those trained with combined 2D–3D visual aids showed the greatest improvement in precision, including reduced tilt and rotation errors, during unaided testing compared to the Control group.
The statistical analysis confirmed that training modality significantly influenced retention, highlighting that training quality matters more than quantity. 
All three modalities were rated highly in terms of usability, usefulness, and ease of use. Notably, the 2D–3D Aid did not increase cognitive workload, suggesting that rich spatial cues can be well-tolerated when aligned with user familiarity.
The execution time was longer with visual aids compared to the Control group. However, the statistical analysis indicated that execution speed improved mainly through repetition, not modality, showing that procedural fluency benefits from practice regardless of the aid used.

\bibliographystyle{unsrtnat}
\bibliography{references}  






\end{document}